\documentclass[twocolumn,prd,amsmath,amssymb,showpacs,nofootinbib,superscriptaddress]{revtex4-1}

\usepackage{color,graphicx}
\usepackage[caption=false]{subfig}

\graphicspath{{figures/}}

\begin{document}

\title{Cosmic-ray electron+positron spectrum from 7~GeV to 2~TeV with the Fermi Large Area Telescope}

\author{S.~Abdollahi}
\affiliation{Department of Physical Sciences, Hiroshima University, Higashi-Hiroshima, Hiroshima 739-8526, Japan}
\author{M.~Ackermann}
\affiliation{Deutsches Elektronen Synchrotron DESY, D-15738 Zeuthen, Germany}
\author{M.~Ajello}
\affiliation{Department of Physics and Astronomy, Clemson University, Kinard Lab of Physics, Clemson, SC 29634-0978, USA}
\author{W.~B.~Atwood}
\affiliation{Santa Cruz Institute for Particle Physics, Department of Physics and Department of Astronomy and Astrophysics, University of California at Santa Cruz, Santa Cruz, CA 95064, USA}
\author{L.~Baldini}
\affiliation{Universit\`a di Pisa and Istituto Nazionale di Fisica Nucleare, Sezione di Pisa I-56127 Pisa, Italy}
\author{G.~Barbiellini}
\affiliation{Istituto Nazionale di Fisica Nucleare, Sezione di Trieste, I-34127 Trieste, Italy}
\affiliation{Dipartimento di Fisica, Universit\`a di Trieste, I-34127 Trieste, Italy}
\author{D.~Bastieri}
\affiliation{Istituto Nazionale di Fisica Nucleare, Sezione di Padova, I-35131 Padova, Italy}
\affiliation{Dipartimento di Fisica e Astronomia ``G. Galilei'', Universit\`a di Padova, I-35131 Padova, Italy}
\author{R.~Bellazzini}
\affiliation{Istituto Nazionale di Fisica Nucleare, Sezione di Pisa, I-56127 Pisa, Italy}
\author{E.~D.~Bloom}
\affiliation{W. W. Hansen Experimental Physics Laboratory, Kavli Institute for Particle Astrophysics and Cosmology, Department of Physics and SLAC National Accelerator Laboratory, Stanford University, Stanford, CA 94305, USA}
\author{R.~Bonino}
\email{rbonino@to.infn.it}
\affiliation{Istituto Nazionale di Fisica Nucleare, Sezione di Torino, I-10125 Torino, Italy}
\affiliation{Dipartimento di Fisica, Universit\`a degli Studi di Torino, I-10125 Torino, Italy}
\author{T.~J.~Brandt}
\affiliation{NASA Goddard Space Flight Center, Greenbelt, MD 20771, USA}
\author{J.~Bregeon}
\affiliation{Laboratoire Univers et Particules de Montpellier, Universit\'e Montpellier, CNRS/IN2P3, F-34095 Montpellier, France}
\author{P.~Bruel}
\email{Philippe.Bruel@llr.in2p3.fr}
\affiliation{Laboratoire Leprince-Ringuet, \'Ecole polytechnique, CNRS/IN2P3, F-91128 Palaiseau, France}
\author{R.~Buehler}
\affiliation{Deutsches Elektronen Synchrotron DESY, D-15738 Zeuthen, Germany}
\author{R.~A.~Cameron}
\affiliation{W. W. Hansen Experimental Physics Laboratory, Kavli Institute for Particle Astrophysics and Cosmology, Department of Physics and SLAC National Accelerator Laboratory, Stanford University, Stanford, CA 94305, USA}
\author{R.~Caputo}
\affiliation{Santa Cruz Institute for Particle Physics, Department of Physics and Department of Astronomy and Astrophysics, University of California at Santa Cruz, Santa Cruz, CA 95064, USA}
\author{M.~Caragiulo}
\affiliation{Dipartimento di Fisica ``M. Merlin" dell'Universit\`a e del Politecnico di Bari, I-70126 Bari, Italy}
\affiliation{Istituto Nazionale di Fisica Nucleare, Sezione di Bari, I-70126 Bari, Italy}
\author{D.~Castro}
\affiliation{NASA Goddard Space Flight Center, Greenbelt, MD 20771, USA}
\author{E.~Cavazzuti}
\affiliation{Agenzia Spaziale Italiana (ASI) Science Data Center, I-00133 Roma, Italy}
\author{C.~Cecchi}
\affiliation{Istituto Nazionale di Fisica Nucleare, Sezione di Perugia, I-06123 Perugia, Italy}
\affiliation{Dipartimento di Fisica, Universit\`a degli Studi di Perugia, I-06123 Perugia, Italy}
\author{A.~Chekhtman}
\affiliation{College of Science, George Mason University, Fairfax, VA 22030, resident at Naval Research Laboratory, Washington, DC 20375, USA}
\author{S.~Ciprini}
\affiliation{Agenzia Spaziale Italiana (ASI) Science Data Center, I-00133 Roma, Italy}
\affiliation{Istituto Nazionale di Fisica Nucleare, Sezione di Perugia, I-06123 Perugia, Italy}
\author{J.~Cohen-Tanugi}
\affiliation{Laboratoire Univers et Particules de Montpellier, Universit\'e Montpellier, CNRS/IN2P3, F-34095 Montpellier, France}
\author{F.~Costanza}
\affiliation{Istituto Nazionale di Fisica Nucleare, Sezione di Bari, I-70126 Bari, Italy}
\author{A.~Cuoco}
\affiliation{RWTH Aachen University, Institute for Theoretical Particle Physics and Cosmology, (TTK),, D-52056 Aachen, Germany}
\affiliation{Istituto Nazionale di Fisica Nucleare, Sezione di Torino, I-10125 Torino, Italy}
\author{S.~Cutini}
\affiliation{Agenzia Spaziale Italiana (ASI) Science Data Center, I-00133 Roma, Italy}
\affiliation{Istituto Nazionale di Fisica Nucleare, Sezione di Perugia, I-06123 Perugia, Italy}
\author{F.~D'Ammando}
\affiliation{INAF Istituto di Radioastronomia, I-40129 Bologna, Italy}
\affiliation{Dipartimento di Astronomia, Universit\`a di Bologna, I-40127 Bologna, Italy}
\author{F.~de~Palma}
\affiliation{Istituto Nazionale di Fisica Nucleare, Sezione di Bari, I-70126 Bari, Italy}
\affiliation{Universit\`a Telematica Pegaso, Piazza Trieste e Trento, 48, I-80132 Napoli, Italy}
\author{R.~Desiante}
\affiliation{Istituto Nazionale di Fisica Nucleare, Sezione di Torino, I-10125 Torino, Italy}
\affiliation{Universit\`a di Udine, I-33100 Udine, Italy}
\author{S.~W.~Digel}
\affiliation{W. W. Hansen Experimental Physics Laboratory, Kavli Institute for Particle Astrophysics and Cosmology, Department of Physics and SLAC National Accelerator Laboratory, Stanford University, Stanford, CA 94305, USA}
\author{N.~Di~Lalla}
\affiliation{Istituto Nazionale di Fisica Nucleare, Sezione di Pisa, I-56127 Pisa, Italy}
\author{M.~Di~Mauro}
\affiliation{W. W. Hansen Experimental Physics Laboratory, Kavli Institute for Particle Astrophysics and Cosmology, Department of Physics and SLAC National Accelerator Laboratory, Stanford University, Stanford, CA 94305, USA}
\author{L.~Di~Venere}
\affiliation{Dipartimento di Fisica ``M. Merlin" dell'Universit\`a e del Politecnico di Bari, I-70126 Bari, Italy}
\affiliation{Istituto Nazionale di Fisica Nucleare, Sezione di Bari, I-70126 Bari, Italy}
\author{P.~S.~Drell}
\affiliation{W. W. Hansen Experimental Physics Laboratory, Kavli Institute for Particle Astrophysics and Cosmology, Department of Physics and SLAC National Accelerator Laboratory, Stanford University, Stanford, CA 94305, USA}
\author{A.~Drlica-Wagner}
\affiliation{Center for Particle Astrophysics, Fermi National Accelerator Laboratory, Batavia, IL 60510, USA}
\author{C.~Favuzzi}
\affiliation{Dipartimento di Fisica ``M. Merlin" dell'Universit\`a e del Politecnico di Bari, I-70126 Bari, Italy}
\affiliation{Istituto Nazionale di Fisica Nucleare, Sezione di Bari, I-70126 Bari, Italy}
\author{W.~B.~Focke}
\affiliation{W. W. Hansen Experimental Physics Laboratory, Kavli Institute for Particle Astrophysics and Cosmology, Department of Physics and SLAC National Accelerator Laboratory, Stanford University, Stanford, CA 94305, USA}
\author{S.~Funk}
\affiliation{Erlangen Centre for Astroparticle Physics, D-91058 Erlangen, Germany}
\author{P.~Fusco}
\affiliation{Dipartimento di Fisica ``M. Merlin" dell'Universit\`a e del Politecnico di Bari, I-70126 Bari, Italy}
\affiliation{Istituto Nazionale di Fisica Nucleare, Sezione di Bari, I-70126 Bari, Italy}
\author{F.~Gargano}
\affiliation{Istituto Nazionale di Fisica Nucleare, Sezione di Bari, I-70126 Bari, Italy}
\author{D.~Gasparrini}
\affiliation{Agenzia Spaziale Italiana (ASI) Science Data Center, I-00133 Roma, Italy}
\affiliation{Istituto Nazionale di Fisica Nucleare, Sezione di Perugia, I-06123 Perugia, Italy}
\author{N.~Giglietto}
\affiliation{Dipartimento di Fisica ``M. Merlin" dell'Universit\`a e del Politecnico di Bari, I-70126 Bari, Italy}
\affiliation{Istituto Nazionale di Fisica Nucleare, Sezione di Bari, I-70126 Bari, Italy}
\author{F.~Giordano}
\affiliation{Dipartimento di Fisica ``M. Merlin" dell'Universit\`a e del Politecnico di Bari, I-70126 Bari, Italy}
\affiliation{Istituto Nazionale di Fisica Nucleare, Sezione di Bari, I-70126 Bari, Italy}
\author{M.~Giroletti}
\affiliation{INAF Istituto di Radioastronomia, I-40129 Bologna, Italy}
\author{D.~Green}
\affiliation{Department of Physics and Department of Astronomy, University of Maryland, College Park, MD 20742, USA}
\affiliation{NASA Goddard Space Flight Center, Greenbelt, MD 20771, USA}
\author{L.~Guillemot}
\affiliation{Laboratoire de Physique et Chimie de l'Environnement et de l'Espace -- Universit\'e d'Orl\'eans / CNRS, F-45071 Orl\'eans Cedex 02, France}
\affiliation{Station de radioastronomie de Nan\c{c}ay, Observatoire de Paris, CNRS/INSU, F-18330 Nan\c{c}ay, France}
\author{S.~Guiriec}
\affiliation{NASA Goddard Space Flight Center, Greenbelt, MD 20771, USA}
\affiliation{NASA Postdoctoral Program Fellow, USA}
\author{A.~K.~Harding}
\affiliation{NASA Goddard Space Flight Center, Greenbelt, MD 20771, USA}
\author{T.~Jogler}
\affiliation{Friedrich-Alexander-Universit\"at, Erlangen-N\"urnberg, Schlossplatz 4, 91054 Erlangen, Germany}
\author{G.~J\'ohannesson}
\affiliation{Science Institute, University of Iceland, IS-107 Reykjavik, Iceland}
\author{T.~Kamae}
\affiliation{Department of Physics, Graduate School of Science, University of Tokyo, 7-3-1 Hongo, Bunkyo-ku, Tokyo 113-0033, Japan}
\author{M.~Kuss}
\affiliation{Istituto Nazionale di Fisica Nucleare, Sezione di Pisa, I-56127 Pisa, Italy}
\author{G.~La~Mura}
\affiliation{Dipartimento di Fisica e Astronomia ``G. Galilei'', Universit\`a di Padova, I-35131 Padova, Italy}
\author{L.~Latronico}
\affiliation{Istituto Nazionale di Fisica Nucleare, Sezione di Torino, I-10125 Torino, Italy}
\author{F.~Longo}
\affiliation{Istituto Nazionale di Fisica Nucleare, Sezione di Trieste, I-34127 Trieste, Italy}
\affiliation{Dipartimento di Fisica, Universit\`a di Trieste, I-34127 Trieste, Italy}
\author{F.~Loparco}
\affiliation{Dipartimento di Fisica ``M. Merlin" dell'Universit\`a e del Politecnico di Bari, I-70126 Bari, Italy}
\affiliation{Istituto Nazionale di Fisica Nucleare, Sezione di Bari, I-70126 Bari, Italy}
\author{P.~Lubrano}
\affiliation{Istituto Nazionale di Fisica Nucleare, Sezione di Perugia, I-06123 Perugia, Italy}
\author{S.~Maldera}
\affiliation{Istituto Nazionale di Fisica Nucleare, Sezione di Torino, I-10125 Torino, Italy}
\author{D.~Malyshev}
\affiliation{Erlangen Centre for Astroparticle Physics, D-91058 Erlangen, Germany}
\author{A.~Manfreda}
\email{alberto.manfreda@pi.infn.it}
\affiliation{Istituto Nazionale di Fisica Nucleare, Sezione di Pisa, I-56127 Pisa, Italy}
\author{M.~N.~Mazziotta}
\affiliation{Istituto Nazionale di Fisica Nucleare, Sezione di Bari, I-70126 Bari, Italy}
\author{P.~F.~Michelson}
\affiliation{W. W. Hansen Experimental Physics Laboratory, Kavli Institute for Particle Astrophysics and Cosmology, Department of Physics and SLAC National Accelerator Laboratory, Stanford University, Stanford, CA 94305, USA}
\author{N.~Mirabal}
\affiliation{NASA Goddard Space Flight Center, Greenbelt, MD 20771, USA}
\affiliation{NASA Postdoctoral Program Fellow, USA}
\author{W.~Mitthumsiri}
\affiliation{Department of Physics, Faculty of Science, Mahidol University, Bangkok 10400, Thailand}
\author{T.~Mizuno}
\affiliation{Hiroshima Astrophysical Science Center, Hiroshima University, Higashi-Hiroshima, Hiroshima 739-8526, Japan}
\author{A.~A.~Moiseev}
\affiliation{Center for Research and Exploration in Space Science and Technology (CRESST) and NASA Goddard Space Flight Center, Greenbelt, MD 20771, USA}
\affiliation{Department of Physics and Department of Astronomy, University of Maryland, College Park, MD 20742, USA}
\author{M.~E.~Monzani}
\affiliation{W. W. Hansen Experimental Physics Laboratory, Kavli Institute for Particle Astrophysics and Cosmology, Department of Physics and SLAC National Accelerator Laboratory, Stanford University, Stanford, CA 94305, USA}
\author{A.~Morselli}
\affiliation{Istituto Nazionale di Fisica Nucleare, Sezione di Roma ``Tor Vergata", I-00133 Roma, Italy}
\author{I.~V.~Moskalenko}
\affiliation{W. W. Hansen Experimental Physics Laboratory, Kavli Institute for Particle Astrophysics and Cosmology, Department of Physics and SLAC National Accelerator Laboratory, Stanford University, Stanford, CA 94305, USA}
\author{M.~Negro}
\affiliation{Istituto Nazionale di Fisica Nucleare, Sezione di Torino, I-10125 Torino, Italy}
\affiliation{Dipartimento di Fisica, Universit\`a degli Studi di Torino, I-10125 Torino, Italy}
\author{E.~Nuss}
\affiliation{Laboratoire Univers et Particules de Montpellier, Universit\'e Montpellier, CNRS/IN2P3, F-34095 Montpellier, France}
\author{E.~Orlando}
\affiliation{W. W. Hansen Experimental Physics Laboratory, Kavli Institute for Particle Astrophysics and Cosmology, Department of Physics and SLAC National Accelerator Laboratory, Stanford University, Stanford, CA 94305, USA}
\author{D.~Paneque}
\affiliation{Max-Planck-Institut f\"ur Physik, D-80805 M\"unchen, Germany}
\author{J.~S.~Perkins}
\affiliation{NASA Goddard Space Flight Center, Greenbelt, MD 20771, USA}
\author{M.~Pesce-Rollins}
\affiliation{Istituto Nazionale di Fisica Nucleare, Sezione di Pisa, I-56127 Pisa, Italy}
\author{F.~Piron}
\affiliation{Laboratoire Univers et Particules de Montpellier, Universit\'e Montpellier, CNRS/IN2P3, F-34095 Montpellier, France}
\author{G.~Pivato}
\affiliation{Istituto Nazionale di Fisica Nucleare, Sezione di Pisa, I-56127 Pisa, Italy}
\author{T.~A.~Porter}
\affiliation{W. W. Hansen Experimental Physics Laboratory, Kavli Institute for Particle Astrophysics and Cosmology, Department of Physics and SLAC National Accelerator Laboratory, Stanford University, Stanford, CA 94305, USA}
\author{G.~Principe}
\affiliation{Erlangen Centre for Astroparticle Physics, D-91058 Erlangen, Germany}
\author{S.~Rain\`o}
\affiliation{Dipartimento di Fisica ``M. Merlin" dell'Universit\`a e del Politecnico di Bari, I-70126 Bari, Italy}
\affiliation{Istituto Nazionale di Fisica Nucleare, Sezione di Bari, I-70126 Bari, Italy}
\author{R.~Rando}
\affiliation{Istituto Nazionale di Fisica Nucleare, Sezione di Padova, I-35131 Padova, Italy}
\affiliation{Dipartimento di Fisica e Astronomia ``G. Galilei'', Universit\`a di Padova, I-35131 Padova, Italy}
\author{M.~Razzano}
\affiliation{Istituto Nazionale di Fisica Nucleare, Sezione di Pisa, I-56127 Pisa, Italy}
\affiliation{Funded by contract FIRB-2012-RBFR12PM1F from the Italian Ministry of Education, University and Research (MIUR)}
\author{A.~Reimer}
\affiliation{Institut f\"ur Astro- und Teilchenphysik and Institut f\"ur Theoretische Physik, Leopold-Franzens-Universit\"at Innsbruck, A-6020 Innsbruck, Austria}
\affiliation{W. W. Hansen Experimental Physics Laboratory, Kavli Institute for Particle Astrophysics and Cosmology, Department of Physics and SLAC National Accelerator Laboratory, Stanford University, Stanford, CA 94305, USA}
\author{O.~Reimer}
\affiliation{Institut f\"ur Astro- und Teilchenphysik and Institut f\"ur Theoretische Physik, Leopold-Franzens-Universit\"at Innsbruck, A-6020 Innsbruck, Austria}
\affiliation{W. W. Hansen Experimental Physics Laboratory, Kavli Institute for Particle Astrophysics and Cosmology, Department of Physics and SLAC National Accelerator Laboratory, Stanford University, Stanford, CA 94305, USA}
\author{C.~Sgr\`o}
\affiliation{Istituto Nazionale di Fisica Nucleare, Sezione di Pisa, I-56127 Pisa, Italy}
\author{D.~Simone}
\affiliation{Istituto Nazionale di Fisica Nucleare, Sezione di Bari, I-70126 Bari, Italy}
\author{E.~J.~Siskind}
\affiliation{NYCB Real-Time Computing Inc., Lattingtown, NY 11560-1025, USA}
\author{F.~Spada}
\affiliation{Istituto Nazionale di Fisica Nucleare, Sezione di Pisa, I-56127 Pisa, Italy}
\author{G.~Spandre}
\affiliation{Istituto Nazionale di Fisica Nucleare, Sezione di Pisa, I-56127 Pisa, Italy}
\author{P.~Spinelli}
\affiliation{Dipartimento di Fisica ``M. Merlin" dell'Universit\`a e del Politecnico di Bari, I-70126 Bari, Italy}
\affiliation{Istituto Nazionale di Fisica Nucleare, Sezione di Bari, I-70126 Bari, Italy}
\author{H.~Tajima}
\affiliation{Solar-Terrestrial Environment Laboratory, Nagoya University, Nagoya 464-8601, Japan}
\affiliation{W. W. Hansen Experimental Physics Laboratory, Kavli Institute for Particle Astrophysics and Cosmology, Department of Physics and SLAC National Accelerator Laboratory, Stanford University, Stanford, CA 94305, USA}
\author{J.~B.~Thayer}
\affiliation{W. W. Hansen Experimental Physics Laboratory, Kavli Institute for Particle Astrophysics and Cosmology, Department of Physics and SLAC National Accelerator Laboratory, Stanford University, Stanford, CA 94305, USA}
\author{L.~Tibaldo}
\affiliation{Max-Planck-Institut f\"ur Kernphysik, D-69029 Heidelberg, Germany}
\author{D.~F.~Torres}
\affiliation{Institute of Space Sciences (IEEC-CSIC), Campus UAB, E-08193 Barcelona, Spain}
\affiliation{Instituci\'o Catalana de Recerca i Estudis Avan\c{c}ats (ICREA), Barcelona, Spain}
\author{E.~Troja}
\affiliation{NASA Goddard Space Flight Center, Greenbelt, MD 20771, USA}
\affiliation{Department of Physics and Department of Astronomy, University of Maryland, College Park, MD 20742, USA}
\author{M.~Wood}
\affiliation{W. W. Hansen Experimental Physics Laboratory, Kavli Institute for Particle Astrophysics and Cosmology, Department of Physics and SLAC National Accelerator Laboratory, Stanford University, Stanford, CA 94305, USA}
\author{A.~Worley}
\affiliation{Department of Physics and Astronomy, University of Denver, Denver, CO 80208, USA}
\author{G.~Zaharijas}
\affiliation{Istituto Nazionale di Fisica Nucleare, Sezione di Trieste, and Universit\`a di Trieste, I-34127 Trieste, Italy}
\affiliation{Laboratory for Astroparticle Physics, University of Nova Gorica, Vipavska 13, SI-5000 Nova Gorica, Slovenia}
\author{S.~Zimmer}
\affiliation{University of Geneva, Department of Nuclear and Particle Physics, 24 quai Ernest-Ansermet, CH-1211 Geneva 4, Switzerland}
\collaboration{The Fermi-LAT Collaboration}
\noaffiliation

\begin{abstract}
We present a measurement of the cosmic-ray electron+positron spectrum between 7~GeV and 2~TeV performed with almost seven years of data collected with the {\it Fermi} Large Area Telescope.
We find that the spectrum is well fit by a broken power law with a break energy at about 50 GeV.
Above 50~GeV, the spectrum is well described by a single power law with a spectral index of $3.07 \pm 0.02 \; (\text{stat+syst}) \pm 0.04 \; (\text{energy measurement})$. An exponential cutoff lower than 1.8~TeV is excluded at 95\% CL.
\end{abstract}

\pacs{
  98.70.Sa, 
  96.50.sb, 
  95.85.Ry, 
  95.55.Vj 
}

\maketitle

\section{Introduction}

While propagating throughout the Galaxy, high-energy Cosmic-Ray Electrons and positrons (CRE) rapidly lose energy by interacting with the interstellar radiation field through inverse Compton scattering and by synchrotron emission on the Galactic magnetic field. Their diffusion distance is several hundred parsecs at 1~TeV, much shorter than the radial scale of the Galaxy~\cite{Nishimura1979}. Therefore, the shape of the CRE spectrum from $\sim 100$~GeV up to several TeV (as well as the positron fraction~\cite{Pamela,FermiLATpositrons,AMS_PRLsep2014_positronfraction} and CRE anisotropy~\cite{LATanisotropy2010}) can provide evidence for local CRE sources of astrophysical (supernova remnants and pulsar wind nebulae~\cite{Shen1970,Nishimura1980ApJ,Nishimura1997,Aharonian1995,Kobayashi2004,Blasi:2009hv}) or exotic (dark matter~\cite{Cholis2009PhRvD..80l3511C,Cirelli:2008pk,Bergstrom:2009fa}) nature.

Recent measurements by AMS-02~\cite{AMS_PRLnov2014} and {\it Fermi}~\cite{CRE-PRD-LAT} have shown that the CRE spectrum can be fit with a single power law up to $\sim 1$~TeV, with an index of $3.170 \pm 0.008$ and $3.08 \pm 0.05$, respectively. The H.E.S.S.~\cite{HESS2008,HESS2009} measurements gave the first indication of a cutoff at $\sim 2$~TeV. These results can be interpreted as local CRE sources with a spectral cutoff at about this energy~\cite{DiMauro2014}.

The {\it Fermi} Large Area Telescope (LAT)~\cite{LATmission}, while designed to detect gamma rays, is able to collect and identify CREs with a large acceptance by combining information from its three subsystems, a silicon-strip detector based tracker-converter (TKR), an imaging calorimeter (CAL) consisting of 8 layers of CsI crystals and an anti-coincidence detector (ACD) constructed from tiles of plastic scintillator surrounding the TKR and CAL.

Extending the CRE energy measurement beyond 1~TeV with the LAT is challenging because, at such high energy, only $\sim 35\%$ of the shower is on average contained in the CAL and a significant fraction of the CAL crystals along the shower axis are saturated (crystal saturation occurs when more than $\sim 70$~GeV is deposited in one crystal, which occurs for CREs above $\sim 600$~GeV). Thanks to the new Pass~8 event analysis~\cite{Atwood:2013rka}, with improved track and shower reconstruction as well as improved multivariate methods for background suppression, we can achieve a level of background contamination smaller than 25\% and an energy resolution (defined as the half-width of the 68\% containment range) better than 20\% up to 2~TeV.

In this article, we present an updated measurement of the CRE spectrum, using almost seven years of LAT Pass~8 data up to 2~TeV, performing the first direct measurement above 1~TeV.  A search for anisotropies, using the same data and event selection, and a theoretical interpretation of the CRE spectrum are presented in separate publications~\cite{LATanisotropy2017}~\cite{DiMauro2017}.

The paper is structured as follows: the event selection is introduced in section~\ref{sec:eventselection} and further detailed in the two following sections. The energy measurement and the study of systematic uncertainties are described in sections~\ref{sec:energymeasurement}~and~\ref{sec:systematics}, respectively, while the results are presented and discussed in section~\ref{sec:results}.

\section{Event selection}
\label{sec:eventselection}

The LAT on-board gamma filter is designed to reject charged particles but it accepts all events with a deposited energy in the CAL larger than 20~GeV. As in the previous LAT CRE measurement~\cite{CRE-PRD-LAT}, we performed two independent analyses: the High-Energy (HE) analysis above 42~GeV and the Low-Energy (LE) analysis between 7 and 70~GeV. The former selects events passing the on-board gamma filter, whereas for the latter we use an unbiased sample of all trigger types, prescaled on-board by a factor of 250. In both analyses, we first apply a set of simple cuts before performing a multivariate analysis in order to reduce as much as possible the residual proton contamination.

We use LAT Pass~8 data collected between August~4, 2008 and June~24, 2015, requiring that the rocking angle of the LAT from the zenith is less than 51 degrees. The overall live time for this dataset is 4.68 years. We select events within 60 degrees from the LAT boresight in order to eliminate potential contamination from photons produced in cosmic-ray induced air showers in Earth's atmosphere. The so-called photon Earth limb is located at 113 degrees from the zenith.

We require events with a well reconstructed track, whose path lengths through the CAL are larger than 8 radiation lengths. The event energy is estimated by fitting the shower profile in the CAL. We reject badly reconstructed events by requiring that the $\chi^2/\mathrm{n.d.f.}$ of the shower profile fit is less than 20. 
In order to remove alphas and heavier ions we use the path-length-corrected signal in the ACD and the TKR time over threshold which provide charge-deposition information. Both variables are sensitive to the ionization signal which is proportional to $Z^2$. Applying a cut in the plane spanned by these two variables around the $Z=1$ group reduces contamination of alphas and ions to less than a few per mil with respect to protons. This cut also reduces the residual contribution from celestial photons below $1\%$ of the CRE flux. The LE selection requires more than 2~GeV deposited in the CAL.

For both the HE and the LE analyses, these cuts (hereafter referred to as pre-cuts) are combined with further selections based on multivariate classification analyses. We use the multivariate analysis toolkit TMVA~\cite{TMVA} to train Boosted Decision Trees (BDT) with simulated datasets. After training, each BDT provides a variable $p_\text{BDT}$ between~$-1$ and~$+1$, corresponding to most proton-like and most electron-like events, respectively. Since the $p_\text{BDT}$ distribution for electrons peaks very sharply at~1, we use $P_\text{CRE} = \log_{10}({1-p_\text{BDT}})$ as the CRE estimator. For each energy bin of the analyses, we fit the distribution of $P_\text{CRE}$ with two templates corresponding to simulated electrons and background, respectively. After choosing the cut on $P_\text{CRE}$ that selects the CRE candidates, we use the result of the template fit to estimate and subtract the number of residual background events. The HE and LE multivariate analyses are described in the two following sections.

The simulated datasets that are used to train the BDTs were generated with the standard LAT Monte Carlo (MC) suite~\cite{LATmission}, a detailed simulation of the passage of particles through the LAT based on the Geant4 package~\cite{GEANT4}. Independent simulated datasets were produced to perform data/MC comparisons, model the CRE acceptance and estimate the residual background contamination. Above 42~GeV, the residual background after pre-cuts due to non-proton particles is negligible compared to that due to protons, so the Monte Carlo background sample used in the HE analysis is the output of a pure proton simulation, from 4~GeV to 20~TeV. For the LE analysis, the background MC sample used for the training is a simulation of cosmic rays of both primary and secondary origin in low Earth orbit and Earth limb photons~\cite{Mizuno:2004jb}.

\section{High-Energy analysis}
\label{sec:HEanalysis}

To account for the rapid changes in event topology in the LAT between several GeV and several TeV, we define 8 bins in measured energy (equally spaced in $\log_{10}E$ between 31.6~GeV and 3.16~TeV) and train a BDT for each bin. All BDTs are trained with the same set of variables among the hundreds computed during the LAT event reconstruction. These variables were proven to be the most efficient ones in discriminating between electrons and protons thanks to an optimization procedure using only MC datasets.

These discriminating variables characterize the shower trajectory and topology using information from the CAL and TKR subsystems.
We use TKR-only information, such as the average time over threshold and the number of hits in the three sections of the TKR, as well as CAL-only information, such as estimates of the shower transverse size, the crystal-based $\chi^2$ of the shower profile fit, and the ratio of the energies deposited in the first and second CAL layers. We also use TKR-CAL information like the ratio of the number of TKR hits to the energy deposited in the first two CAL layers and the distance of closest approach of the CAL cluster centroid to the track.

\subsection{Data/MC agreement}
\label{sec:LEanalysis}

The agreement between data and simulation for $P_\text{CRE}$ is critical to the analysis: it ensures the goodness of the template fit, from which the background contamination is estimated, and it drives the precision of the selection efficiency predicted by the simulation. As this agreement depends on the data/MC agreement of the individual variables used as input to the BDTs, we performed a systematic comparison of their distributions measured in data and predicted by the simulation for electrons at various energies (8 bins in $\log_{10}E$ between 31.6~GeV and 3.16~TeV) and incidence angles (5 bins in $\cos\theta$ between 0.5 and 1, where $\theta$ is the angle between the event direction and the LAT boresight).

We found that the widths of the distributions are in good agreement (within 15\%), but, for some variables, the position of the peak is shifted. From the differences of peak positions between data and MC, we derived additive corrections that we parametrized as functions of energy and incidence angle, to ensure good data/MC agreement. We refer to these corrections as the Individual Variables Calibration (IVC) corrections. Rather than applying them to the simulation, we apply them to data. Both solutions are equivalent with respect to BDT efficiency (because the IVC corrections are simple shifts) but the latter is more computationally convenient as it does not require retraining the BDTs.

\begin{figure}[!htb]
  \subfloat{\includegraphics[width=.5\linewidth]{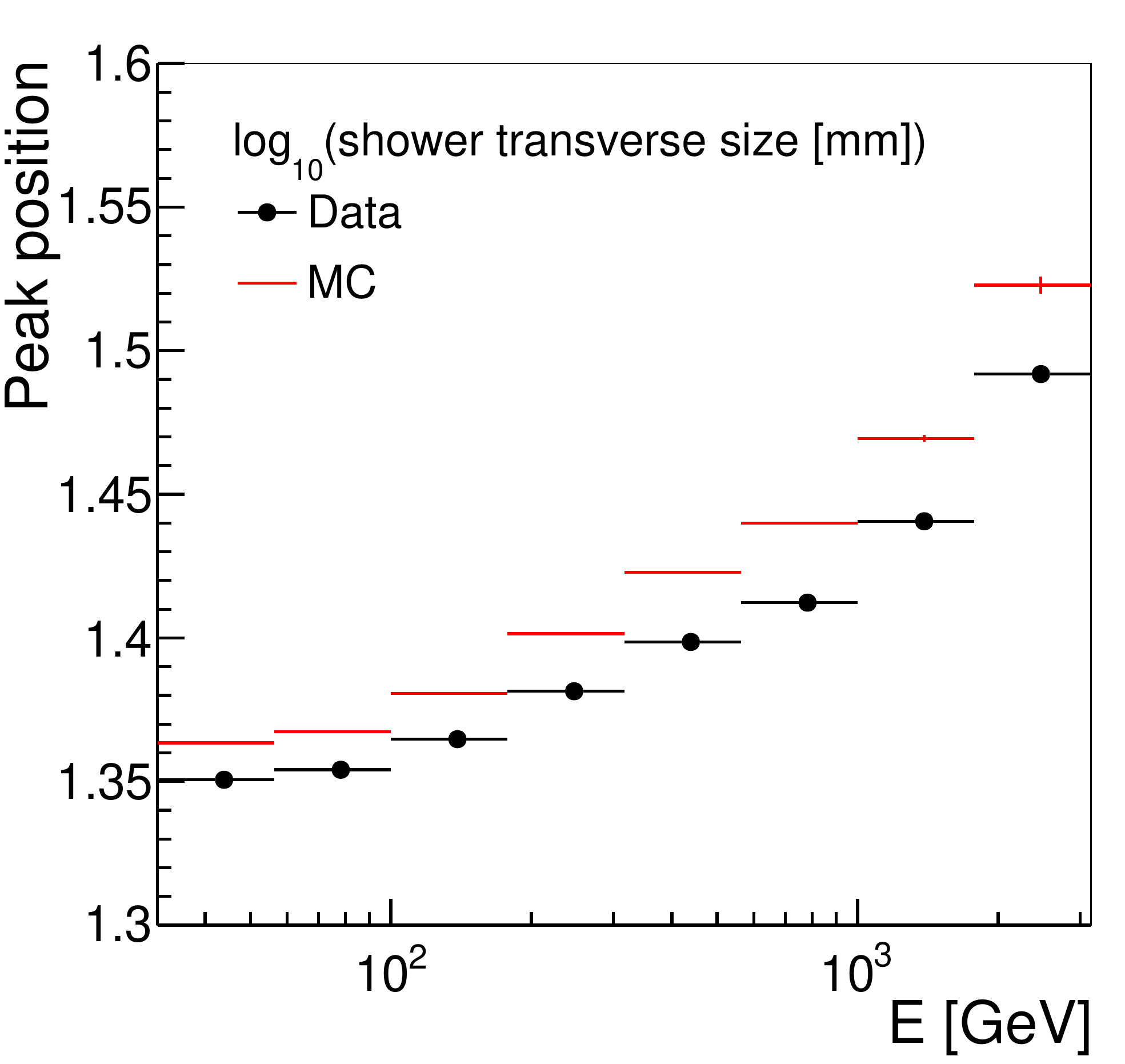}}
  \subfloat{\includegraphics[width=.5\linewidth]{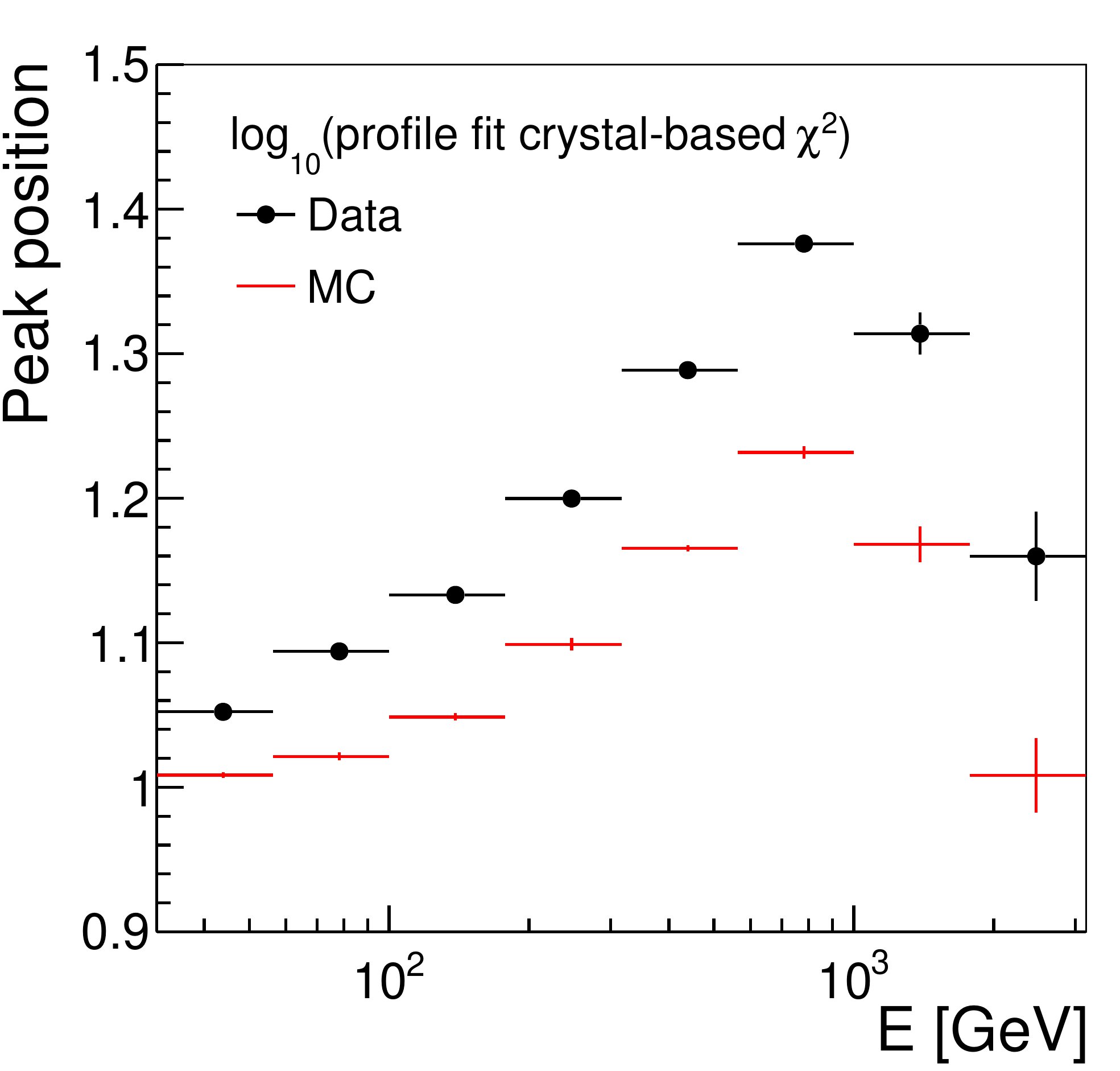}} 
  \caption{Energy dependence of the peak position of the distribution of two variables: the logarithm of the transverse size of the shower (left) and the logarithm of the crystal based $\chi^2$ of the shower profile fit (right). The black and red points correspond to data and simulation respectively.}
  \label{fig:cordat_logEdependence}
\end{figure}

Fig.~\ref{fig:cordat_logEdependence} shows the energy dependence of the peak positions for two variables, the transverse size of the shower and the crystal-based $\chi^2$ of the shower profile fit, for data and MC. Although the data/MC discrepancy for the transverse size of the shower could be solved by rescaling the energy, it is not the case for the crystal-based $\chi^2$ of the shower profile fit. As a consequence, these disagreements are not an indication of a problem in energy measurement. We believe that they are the consequence of imperfections in the instrument simulation.

\begin{figure}[!htb]
  \includegraphics[width=\linewidth]{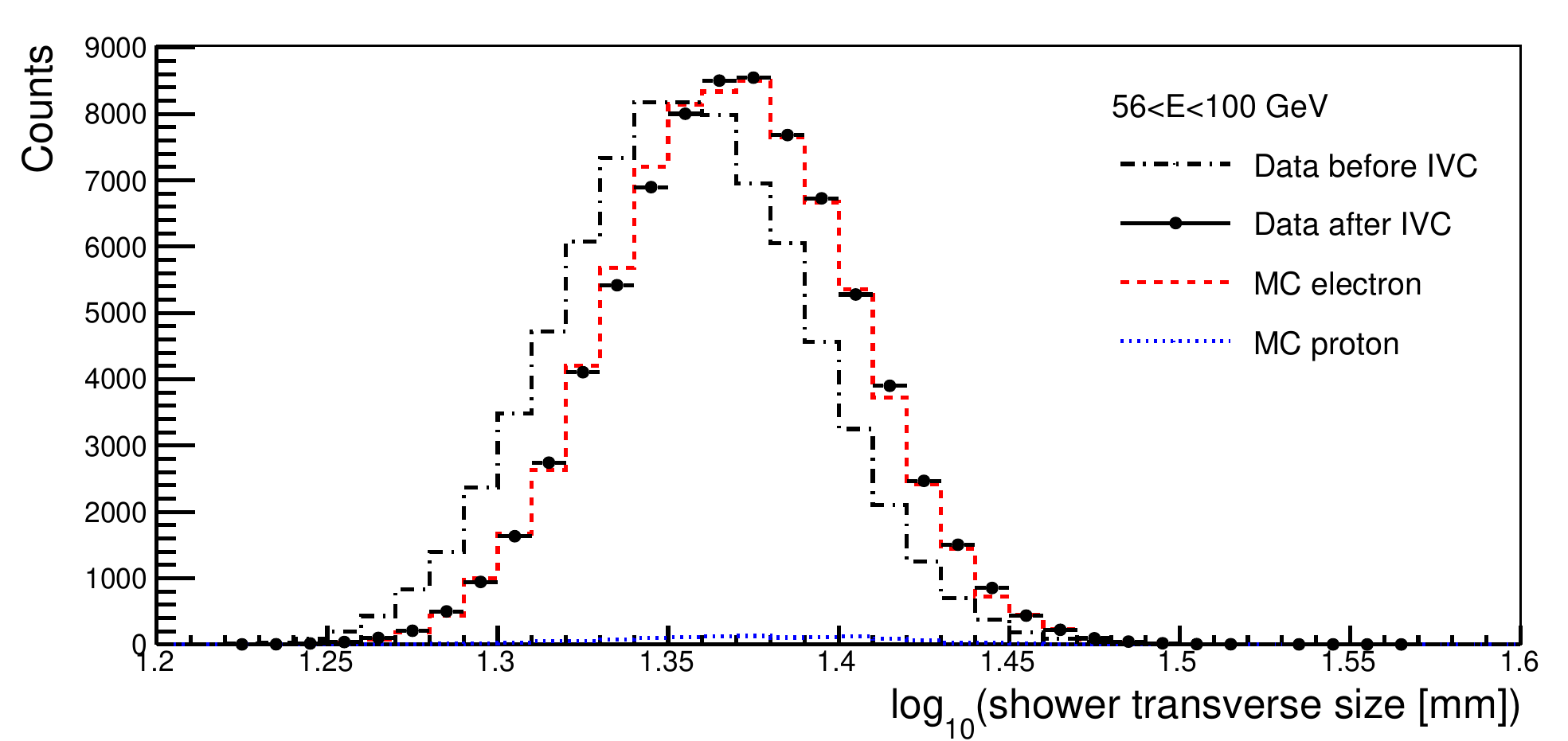}
  \includegraphics[width=\linewidth]{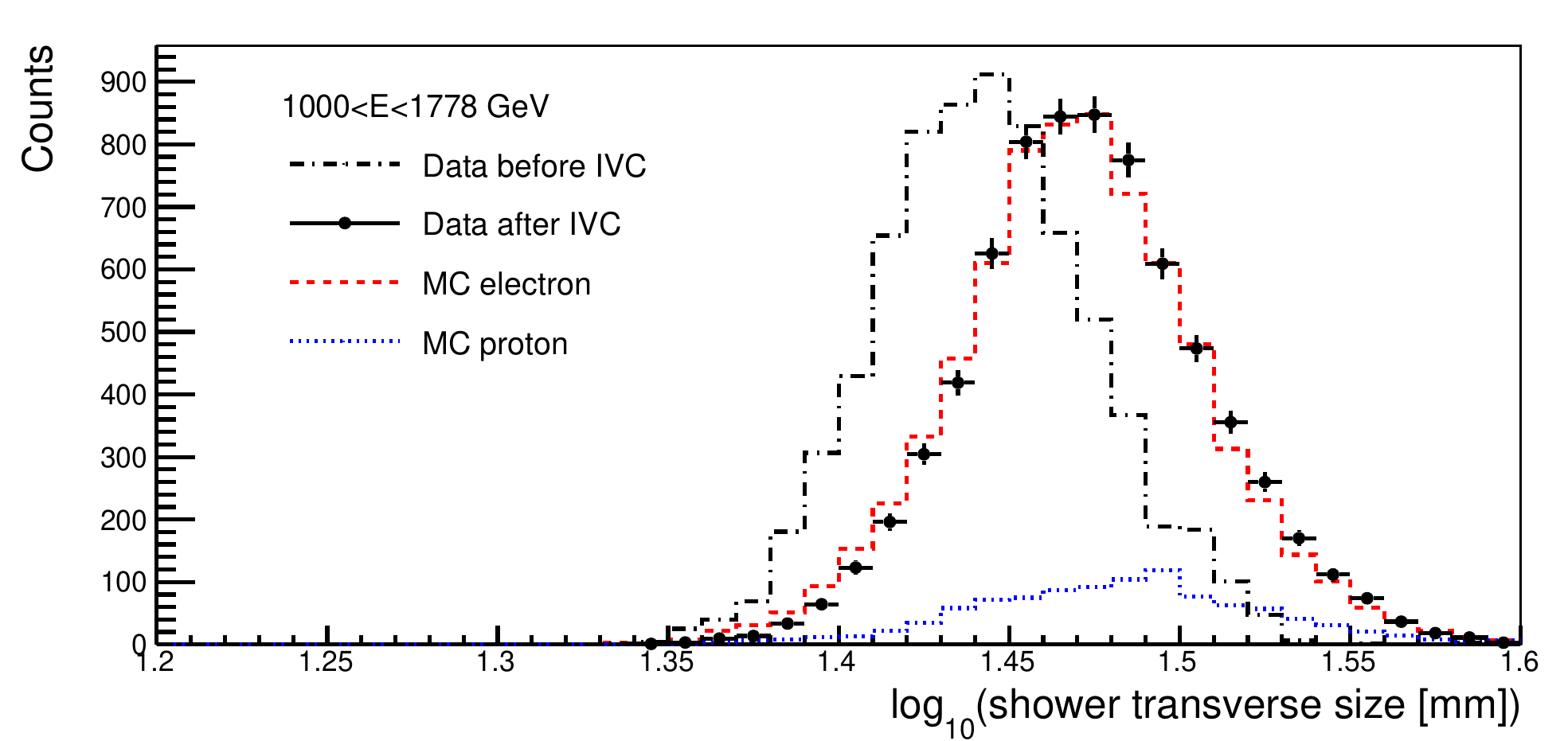}
  \caption{The logarithm of the shower transverse size before (\emph{black lines}) and after (\emph{black circles}) IVC correction for events between 56 and 100~GeV (top) and between 1 and 1.78~TeV (bottom). The contribution from residual background (blue) has been subtracted from the data distributions. The red histograms correspond to the electron simulation.}
  \label{fig:cordat_CTRgeomcor}
\end{figure}

The effect of the IVC corrections is clearly visible in Fig.~\ref{fig:cordat_CTRgeomcor} for the transverse size of the shower, which is one of the variables with the largest data/MC discrepancy. The general trend is that, before correction, the shifts increase with energy and can be as large as the distribution RMS. After correction, the residual differences between peak positions are less than 10\% of the distribution RMS; we take them into account when estimating the systematic uncertainties, as described in section~\ref{sec:systematics}.

\subsection{Template fit}

For each energy bin, we construct the MC electron and proton distributions of $P_\text{CRE}$ and use them as templates: we fit the $P_\text{CRE}$ distribution of the data with the sum of two MC templates whose normalizations are the two parameters of the fit.

Ideally, we would perform the fit over the whole range of $P_\text{CRE}$. Unfortunately, the data/MC agreement after IVC corrections is not good enough in the ``pure'' proton range, {\it i.e.} for $P_\text{CRE}$ above $\sim 0$. This is simply due to the fact that the initial data/MC disagreement of individual variables is not always the same for electrons and protons. The IVC corrections, which are optimized for electrons, cannot ensure a good data/MC agreement for protons. As a result, the peak of the proton $P_\text{CRE}$ distribution ($P_\text{CRE}\gtrsim 0$) is not well reproduced by the simulation.

To mitigate the effect of the data/MC discrepancy near the proton peak we restrict the template fit to an interval $P_\text{CRE}<P_\text{max}$.  To find the optimal value of $P_\text{max}$ we looked at how the $\chi^2$ of the template fit depends on it. We performed a scan starting at $P_\text{max} = -1$, which is well outside the proton peak, but still above the predicted position of the maximum of the signal distribution, and then progressively increasing it. We found that the $\chi^2/\mathrm{n.d.f.}$ remains flat until the fit interval starts to comprise a significant part of the proton peak and we chose $P_\text{max}$ for which the $\chi^2$ has doubled compared to its initial plateau.
Because the data/MC disagreement is larger at high energy, the chosen value of $P_\text{max}$ decreases with energy, as can be seen in Fig.~\ref{fig:bdtcut}.

\begin{figure}[!htb]
  \includegraphics[width=\linewidth]{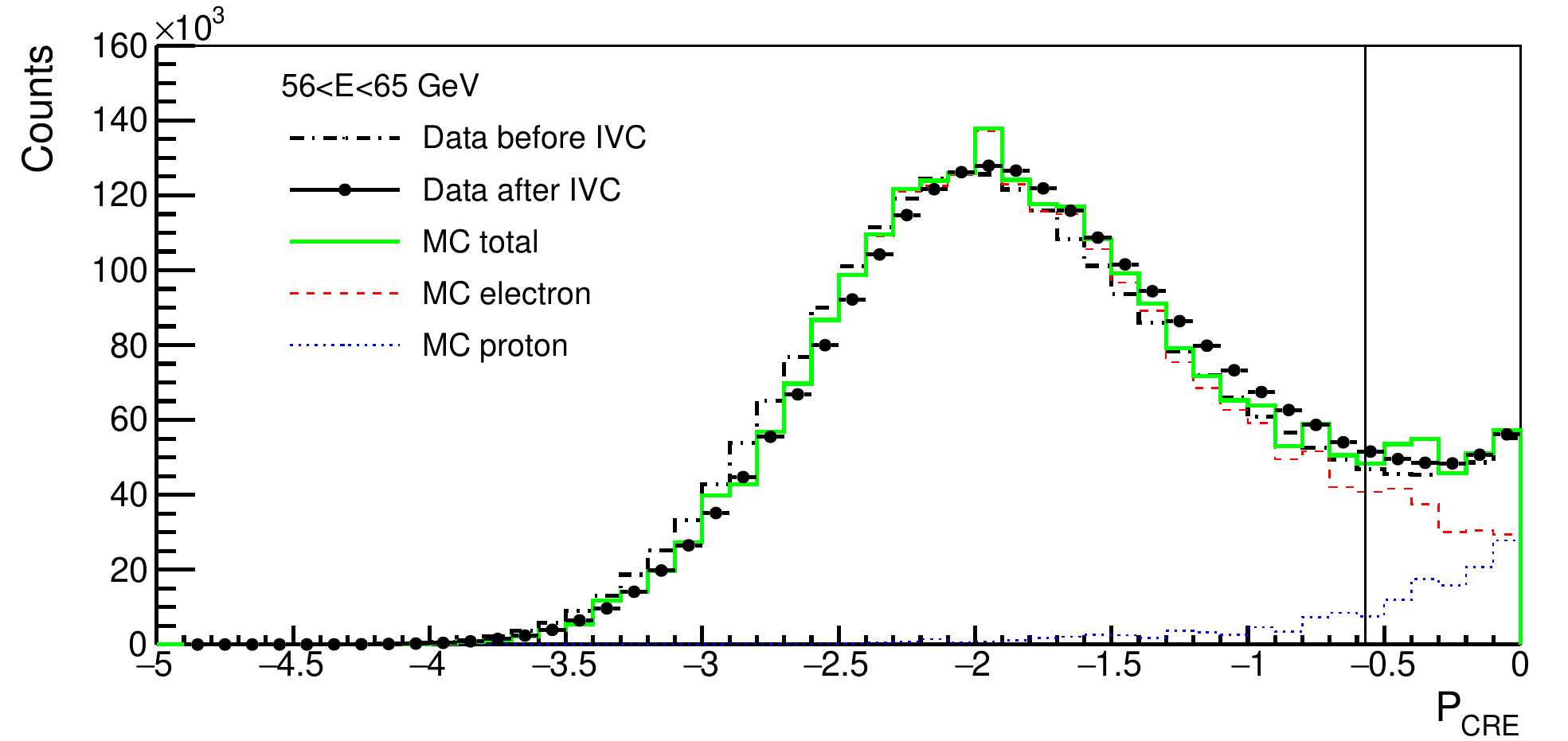}
  \includegraphics[width=\linewidth]{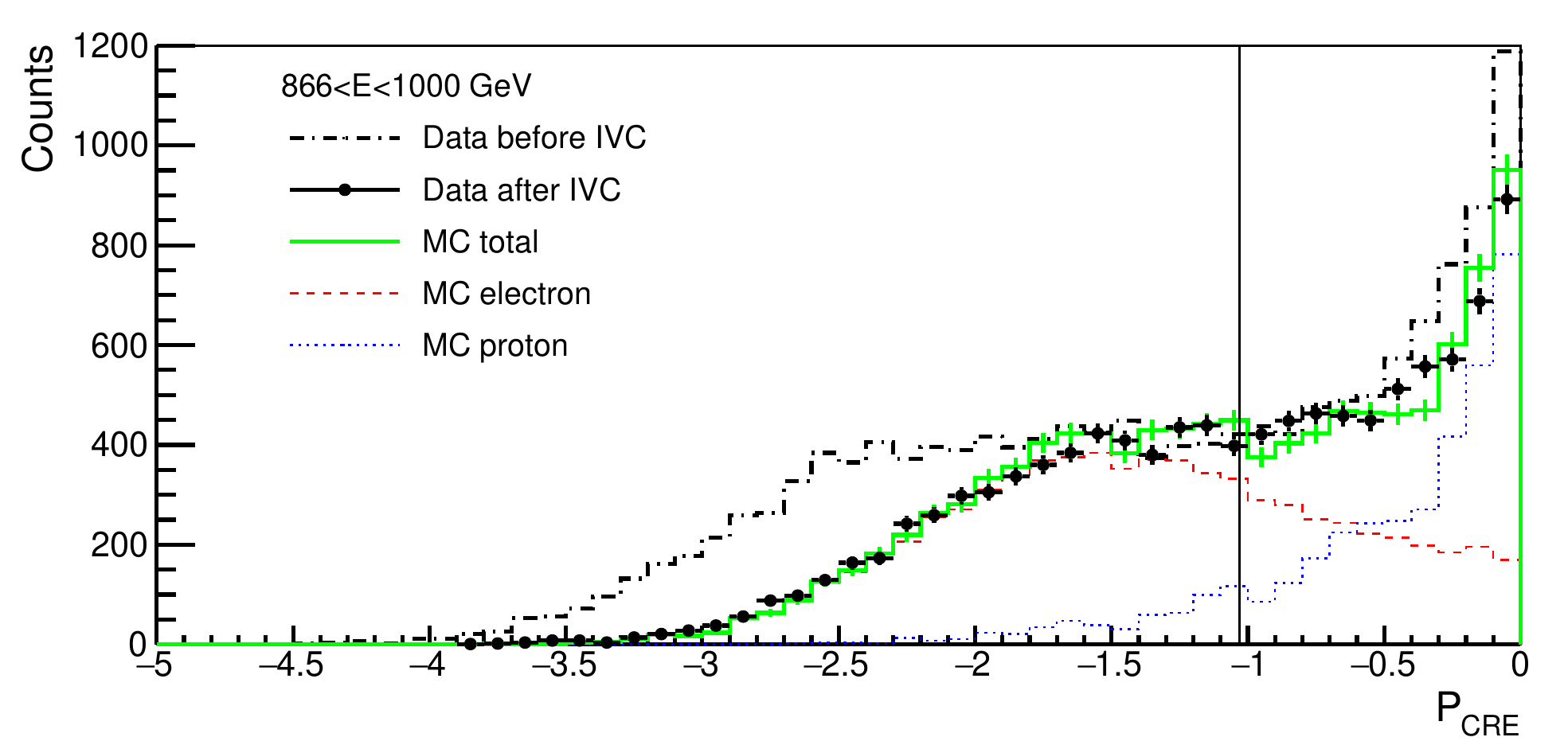}
  \caption{The result of the template fit in two energy bins [56, 65~GeV] (top) and [866, 1000~GeV] (bottom). For both energy bins, the fit is performed over the interval [-5,0] and the x-axis range is chosen to focus on the electron peak region. The data $P_\text{CRE}$ distribution is shown before (\emph{black lines}) and after (\emph{black circles}) IVC corrections. The green histograms correspond to the sum of the electron (red) and proton (blue) templates.  The vertical line show for each energy bin the position of the selection cut $P_\text{CRE}<P_\text{cut}$.}
  \label{fig:cordattemplatefit_sig}
\end{figure}

\begin{figure}[!htb]
  \subfloat{\includegraphics[width=.5\linewidth]{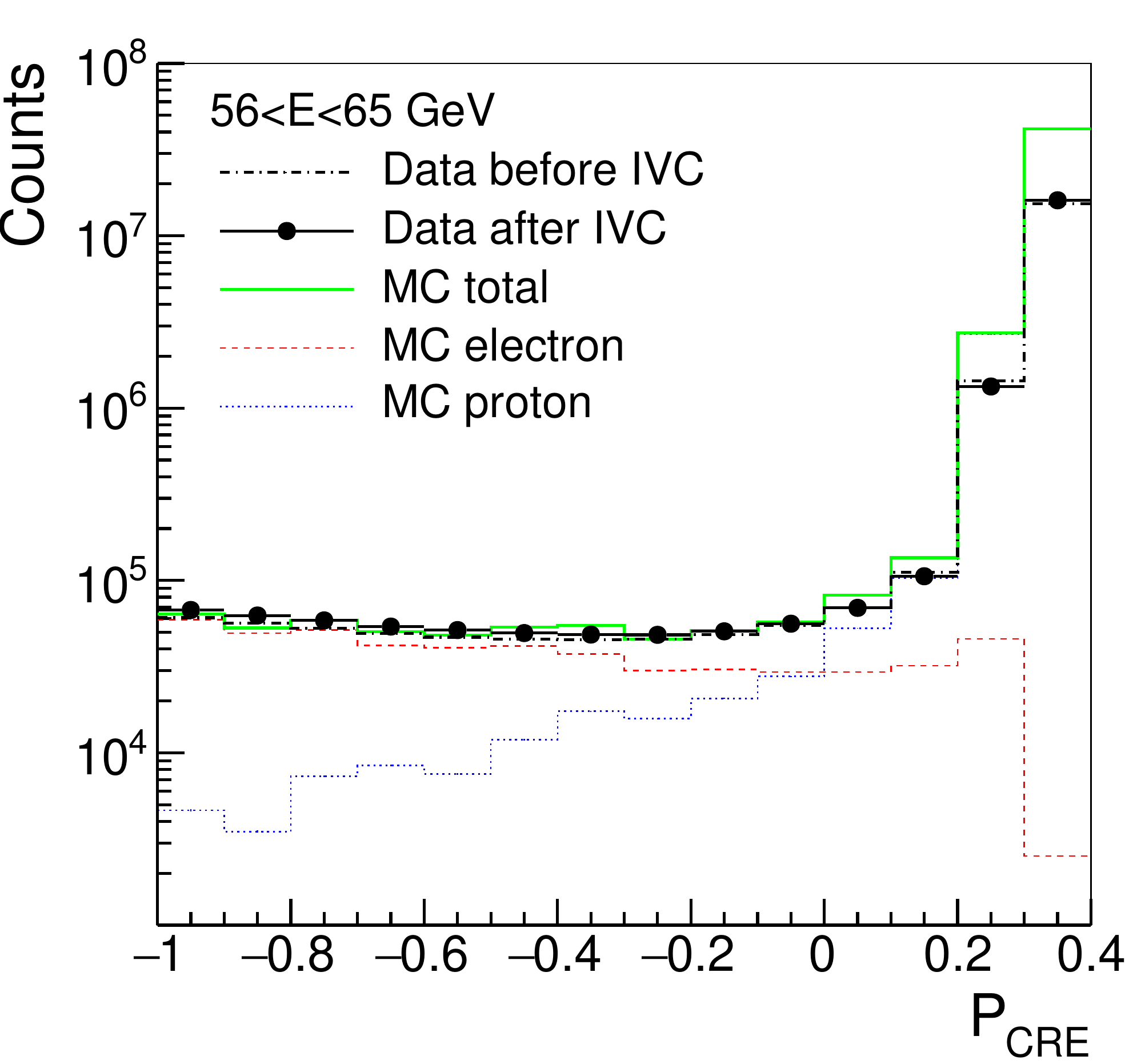}}
  \subfloat{\includegraphics[width=.5\linewidth]{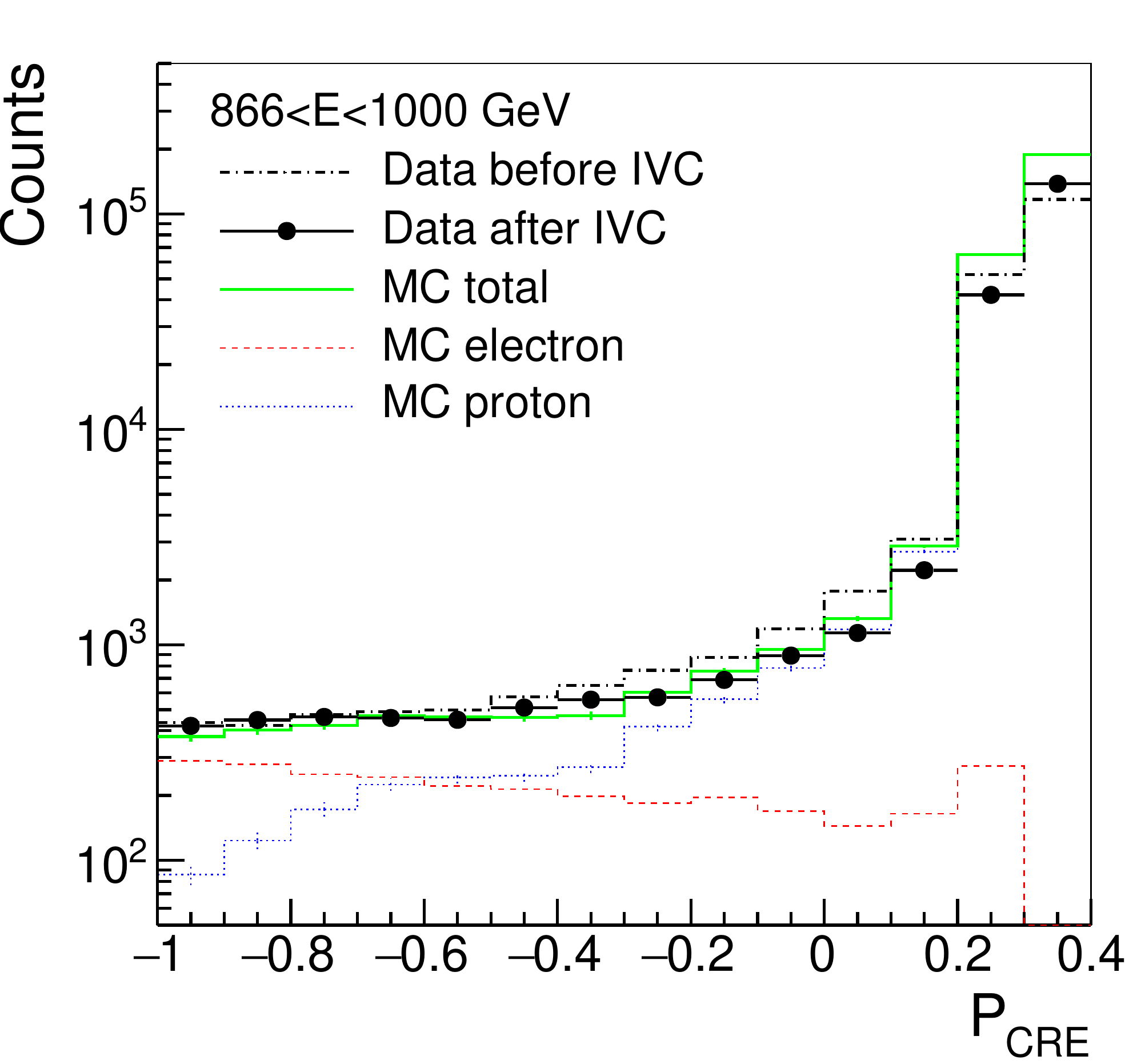}} 
  \caption{The result of the template fit in two energy bins [56, 65~GeV] (left) and [866, 1000~GeV] (right).  For both energy bins, the fit is performed over the interval [-5,0] and the x-axis range is chosen to focus on the proton peak region. The data $P_\text{CRE}$ distribution is shown before (\emph{black lines}) and after (\emph{black circles}) IVC corrections. The green histograms correspond to the sum of the electron (red) and proton (blue) templates.}
  \label{fig:cordattemplatefit_bkg}
\end{figure}

Fig.~\ref{fig:cordattemplatefit_sig}~and~\ref{fig:cordattemplatefit_bkg} show the result of the template fit in two energy bins. One can see that the IVC corrections improve the data/MC agreement for the electron peak but not for the proton peak.

It is to be noted that the data/MC disagreement for the proton peak of the $P_\text{CRE}$ distribution is not an issue for the CRE analysis: the residual background contamination corresponds to protons whose showers appear very much like electromagnetic showers. So, by construction, the IVC corrections are valid for these events and we expect the tail of the proton $P_\text{CRE}$ distribution in the signal region to be well reproduced by the simulation.

\subsection{High-energy selection}

\begin{figure}[!htb]
  \subfloat{\includegraphics[width=.5\linewidth]{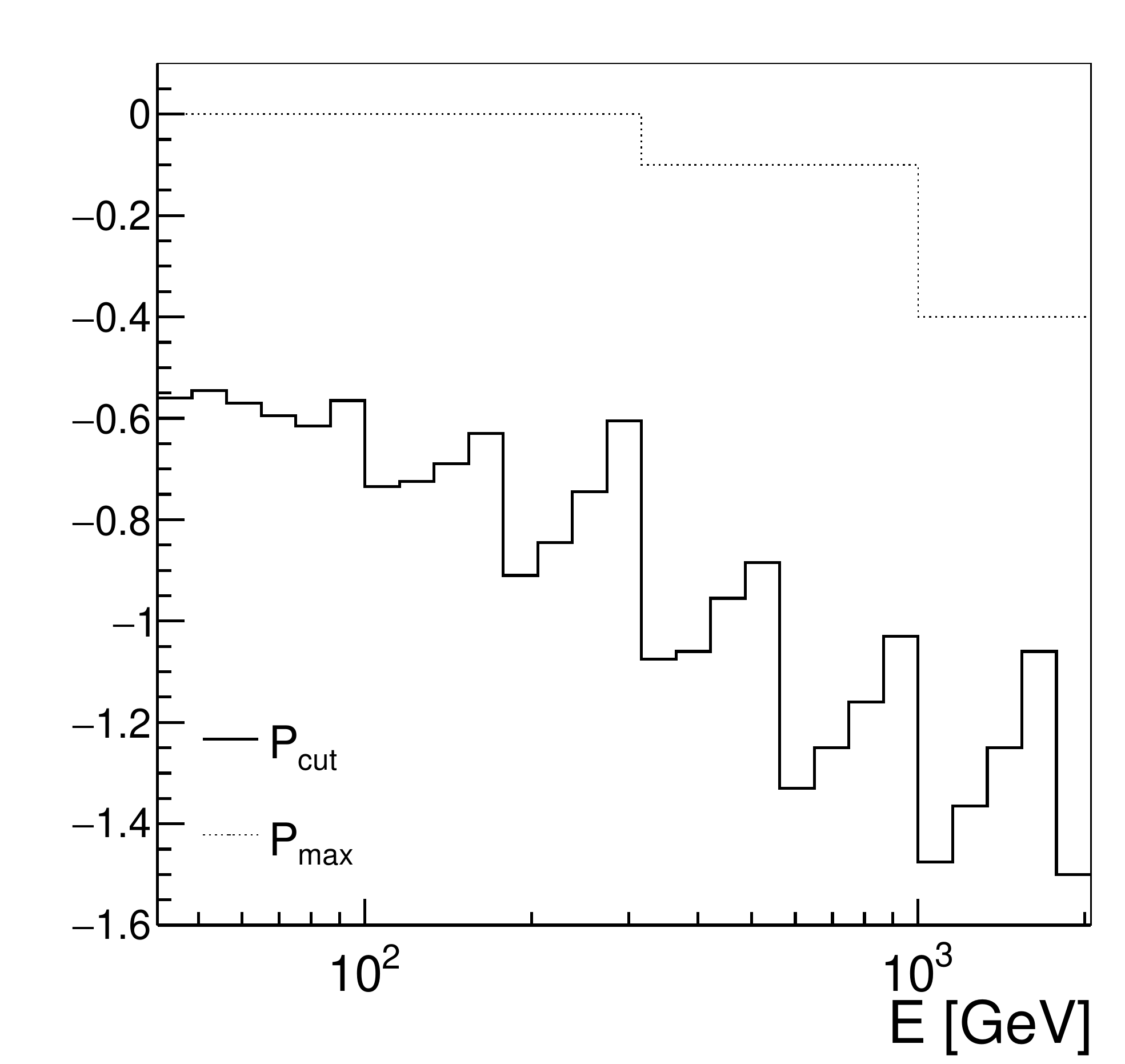}}
  \subfloat{\includegraphics[width=.5\linewidth]{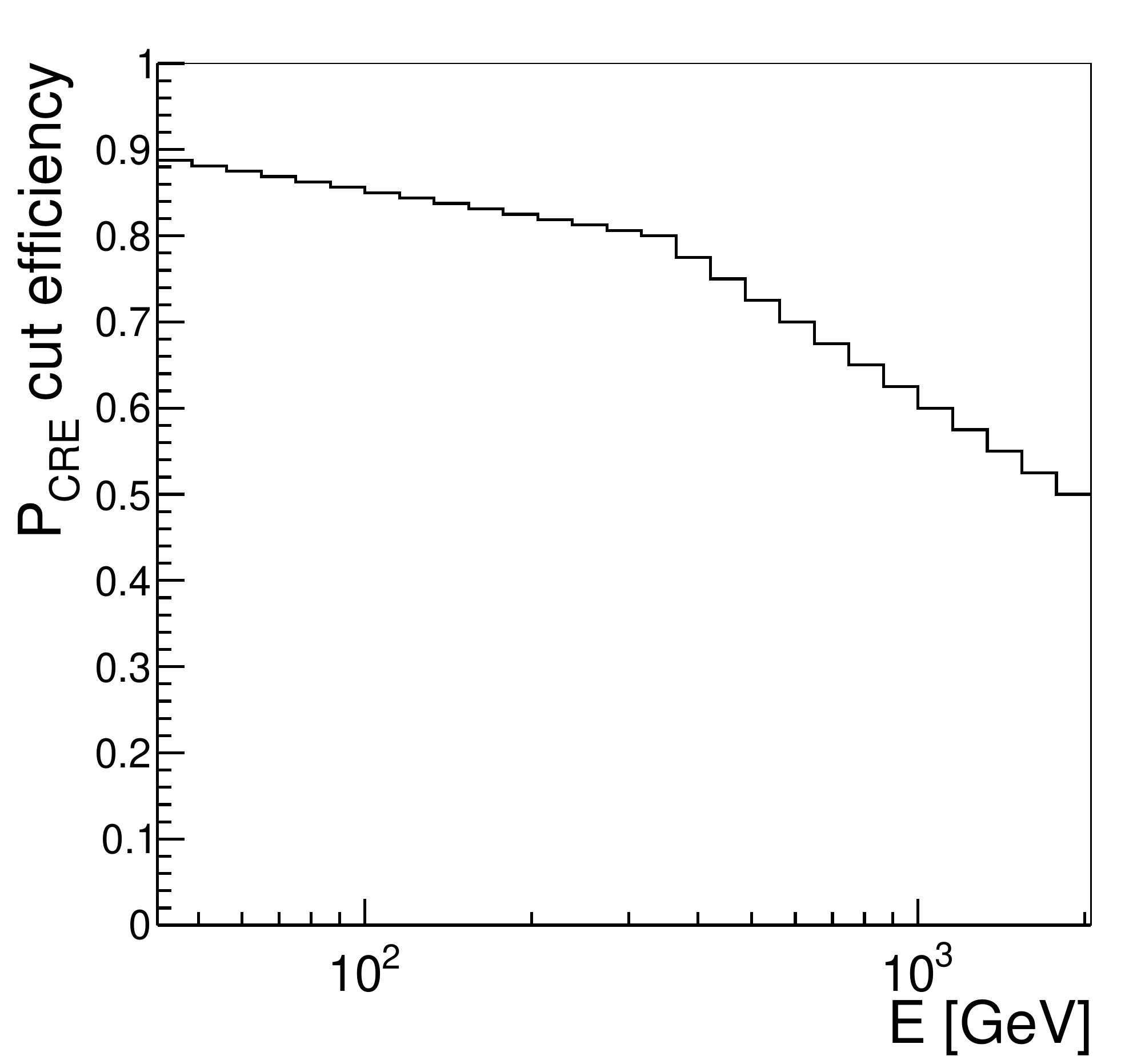}} 
  \caption{Left: $P_\text{CRE}$ selection cut (solid) and $P_\text{max}$, the upper end of the template fit interval (dotted), as a function of energy. Right: efficiency of the $P_\text{CRE}$ selection cut as a function of energy.}
  \label{fig:bdtcut}
\end{figure}

We define our selection by looking, in each energy bin, for the cut on $P_\text{CRE}$ which minimizes the flux uncertainty, taking into account all systematics (discussed in section~\ref{sec:systematics}). The minimum is not very pronounced, especially above 100~GeV. We choose to apply a slightly harder cut between 300~GeV and 2~TeV, in order to facilitate the IVC corrections, whose precision benefits from a low residual background contamination. Fig.~\ref{fig:bdtcut} shows the $P_\text{CRE}$ selection cut and its efficiency as a function of energy. The steps seen every 4 bins in the selection cut correspond to the transitions between the different $P_\text{CRE}$ estimators at the boundaries of the BDT $\log_{10}E$ bins. The  $P_\text{CRE}$ selection cut is well below $P_\text{max}$.

The acceptance and the residual background contamination (defined as the ratio of the number of residual background events to the total number of events) for the HE selection are shown in Fig.~\ref{fig:HELEperformance}. The fact that the acceptance decreases with energy, while the residual contamination increases, highlights the increasing difficulty of the background rejection with energy. Because the IVC corrections procedure requires enough statistics and at the same time a low background contamination, we stop the HE analysis when the residual contamination reaches 20\%, which occurs at 2~TeV.

\section{Low-Energy analysis}

The LE selection is based on the same multivariate analysis approach as used for the HE selection. Because the energy range of the LE selection is much smaller than the HE one, we trained only one BDT, with a set of variables optimized for the LE energy range.
Since the LE analysis stops at 70~GeV, there is no need to apply the IVC corrections.

\begin{figure}[ht]
  \includegraphics[width=.8\linewidth]{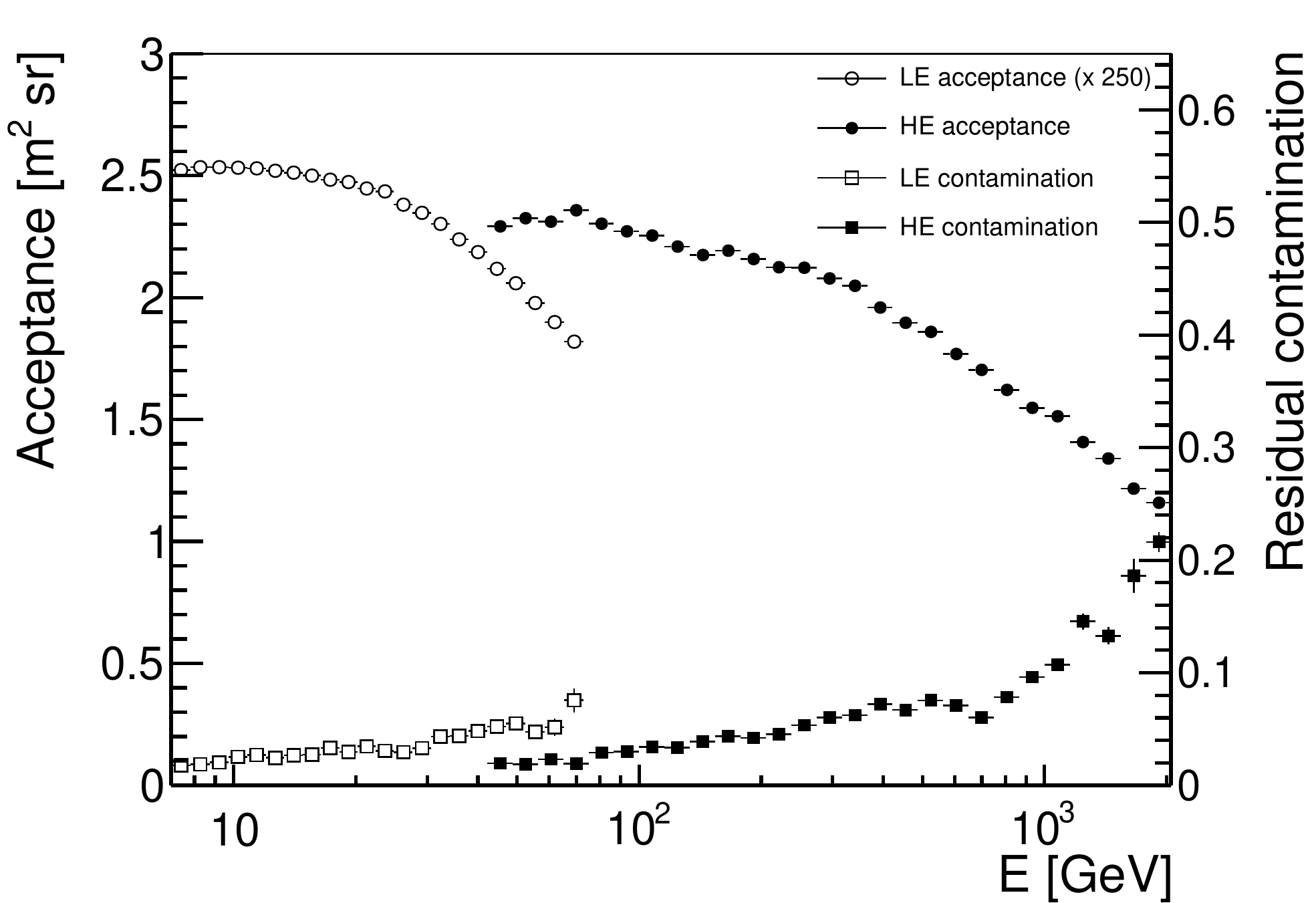}
  \caption{Acceptance and residual background contamination as a function of energy. The displayed LE acceptance is multiplied by 250 (as if there were no prescale factor due to the on-board filter).}
\label{fig:HELEperformance}
\end{figure}

The cut on $P_\text{CRE}$ as a function of energy is set so that the cut efficiency for electrons is $\sim 90\%$ from 7~GeV to 20~GeV and decreases to $\sim 65\%$ at 70~GeV. The residual background contamination is estimated by the same template fitting technique used in the HE analysis. The acceptance and the residual background contamination for the LE selection are shown in Fig.~\ref{fig:HELEperformance}.

Below $\sim 20$~GeV, the flux of CREs observed by the LAT is strongly influenced by the shielding effect of the magnetic field of the Earth. At a given geomagnetic position and direction with respect to zenith, Galactic charged particles can reach the detector only if they are above a certain rigidity. The dependence of the rigidity cutoff on location can be conveniently parametrized by the McIlwain~L parameter~\cite{SmartShea}. Geographic coordinates with the same McIlwain~L parameter are magnetically equivalent from the standpoint of incoming charged particles. The orbit of {\it Fermi} spans an interval of McIlwain~L of $0.98$--$1.73$, corresponding to vertical rigidity cutoff values from $\sim 6$~GeV to $\sim 14$~GeV. Therefore, measuring the CRE spectrum at a given energy E requires selecting data collected in a McIlwain~L interval in which the rigidity cutoff is smaller than E.

\begin{figure}[!htb]
\includegraphics[width=.8\linewidth]{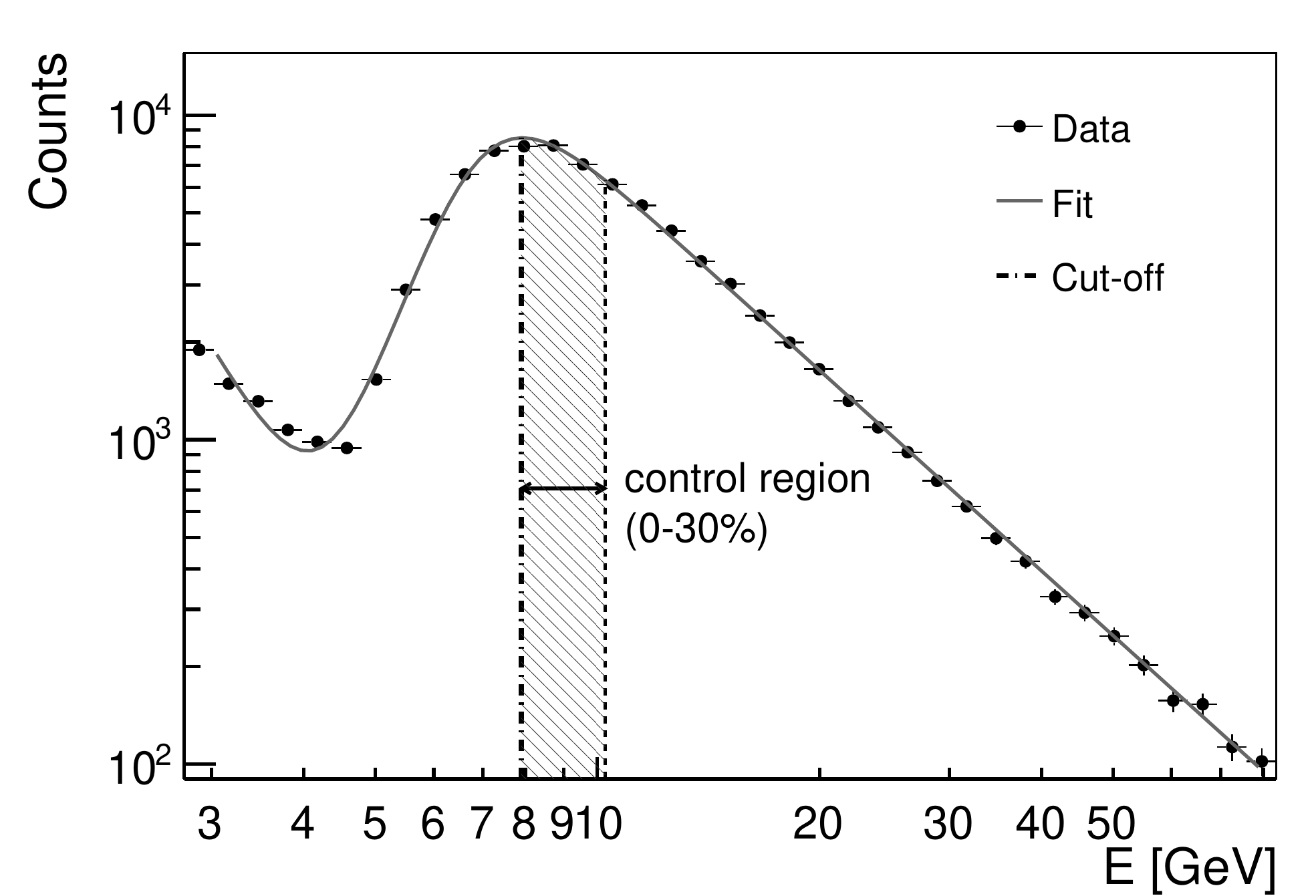}
\caption{Fit of the count spectrum for the McIlwain L bin [1.44, 1.49]. The control region, used in the estimation of the systematic uncertainties, corresponds to the interval $[E_c,1.3E_c]$, where $E_c$ is the measured energy cutoff.}
\label{fig:LEcutoff}
\end{figure}

In order to parametrize the relation between the rigidity cutoff and McIlwain~L, we fit the count spectrum in $15$ McIlwain~L bins with $f(E) = c_{s} E^{{-\gamma}_{s}} + c_{p} E^{{-\gamma}_{p}}/ (1+(E/E_0)^{-\alpha})$. The first term in $f(E)$ corresponds to secondary CREs while the second term in $f(E)$ corresponds to primary CREs~\footnote{In this section we refer to primary CREs as electrons of Galactic origin, and to secondary CREs as splash and reentrant electrons produced in the interactions of primaries in the Earth's atmosphere.}, which are suppressed below the energy $E_0$. An example of this fit is shown in Fig.~\ref{fig:LEcutoff}. We use the position of the maximum of the second term as a measure of the local value of the energy cutoff $E_c$ averaged over the instrument acceptance.

\begin{figure}[!htb]
\includegraphics[width=.8\linewidth]{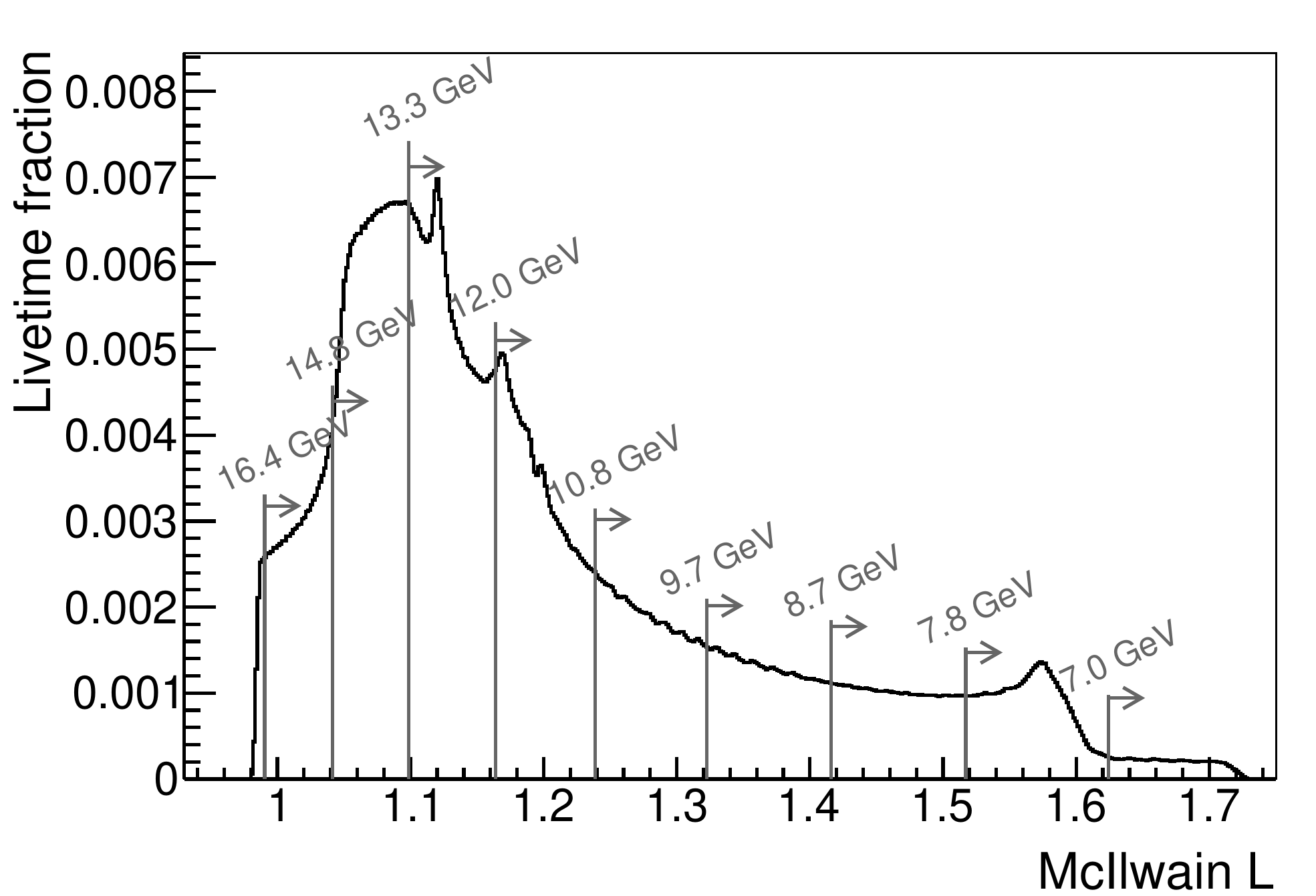}
\caption{Normalized distribution of the instrument live time in different McIlwain~L bins for the data sample. Grey lines show the McIlwain~L cut in each energy bin (with numbers indicating the lower edge of the bin).}
\label{fig:McIlwainLcut}
\end{figure}

We map the dependence of the energy cutoff on the McIlwain~L parameter with the empirical relation $E_c = -14.91 + 67.25/\mathrm{L} -75.71/\mathrm{L}^2 + 39.44/\mathrm{L}^3$. For each energy bin of the CRE spectrum, we use this relation to find the McIlwain~L value $\mathrm{L}_{min}$ corresponding to the lower boundary of the energy bin and select data collected in regions with $\mathrm{L}>\mathrm{L}_{min}$. The corresponding fraction of live time spent by the LAT in the selected regions is $\sim 1.25$\% in the lowest energy bin (from $7$~GeV to $7.8$~GeV) and becomes $\sim 100\%$ above $\sim 18.2$~GeV (see Fig.~\ref{fig:McIlwainLcut}).

Even above the nominal rigidity cutoff a fraction of electrons and positrons is prevented from reaching the detector by the magnetic shadow of the Earth. In order to estimate this fraction of undetected CREs, we use the particle trajectory tracing code (hereafter \emph{tracer}) developed by Smart and Shea~\cite{SmartShea} and the 2010 model of the Earth's magnetic field from the International Geomagnetic Reference Field (IGRF)~\cite{IGRF-11}, as we did in~\cite{LAT_absolute_energy_scale}. For efficiency's sake, \emph{tracer} computes the trajectories of test particles in the reverse direction, starting from the spacecraft. The test particles (electrons and positrons) are generated according to a power law with an index of $3.2$ using the abundance ratios measured by AMS-02 for electrons and positrons~\cite{AMS_PRLsep2014_positronfraction}.

We consider test particles with trajectories reaching 20 Earth radii to have escaped the geomagnetic field, thus corresponding to CREs actually observed by the instrument, while trajectories intersecting the Earth's atmosphere correspond to lost particles. For a given McIlwain L selection, we use the \emph{tracer} output to estimate the fraction of the latter.

\begin{figure}[!htb]
\includegraphics[width=.8\linewidth]{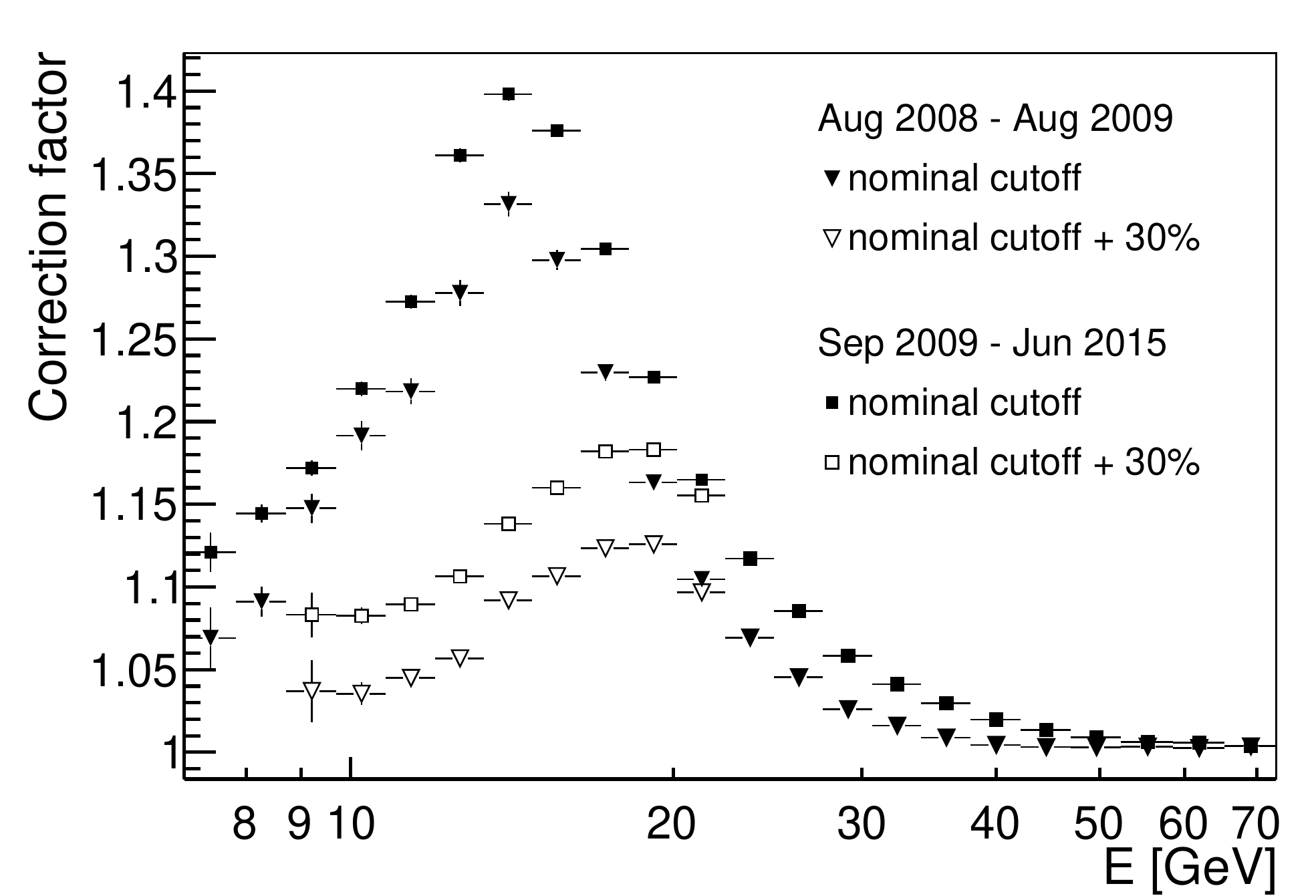}
\caption{LE correction factors for the first year of the mission (triangles) and for the following years (squares). Full markers correspond to the nominal energy cutoff, while the empty markers correspond to a cutoff 30\% higher. Both are almost identical above 22~GeV.}
\label{fig:LEcorrectionfactors}
\end{figure}

The effect of the geomagnetic field on the loss of primaries is enhanced by a combination of the wide angular aperture of the LAT and its periodic rocking motion with respect to the local zenith, with the result that the edge of the field of view is often very close to the Earth. Because of the rocking angle dependence, we derived correction factors separately for the first year of the mission (rocking angle of $35^{\mathrm{o}}$) and for the following years (rocking angle of $50^{\mathrm{o}}$). These correction factors are shown in Fig.~\ref{fig:LEcorrectionfactors}, as well as the correction factors that we obtain when considering energy cutoffs 30\% higher. When estimating the systematic uncertainty for the LE analysis, we vary the energy cutoff choice between $E_c$ and $1.3E_c$. For each energy cutoff, we derive the corresponding McIlwain L selection and \emph{tracer} correction and compute the CRE flux. The CRE fluxes that we obtain are within 3\% of the nominal flux despite the large variation of the \emph{tracer} correction.

\section{Energy measurement}
\label{sec:energymeasurement}

As for the previous versions of the LAT event analysis Passes~6 and~7, the Pass~8 energy reconstruction above $\sim 5$~GeV is performed by fitting the longitudinal shower profile, using the 8 CAL layer energies. The fit parameters are the energy and two parameters that describe the shape of the profile: the shape parameter $\alpha$ and the position of the shower maximum $T_\mathrm{max}$. Further details on the profile fit can be found in Appendix~A.

\begin{figure}[!htb]
  \includegraphics[width=.8\linewidth]{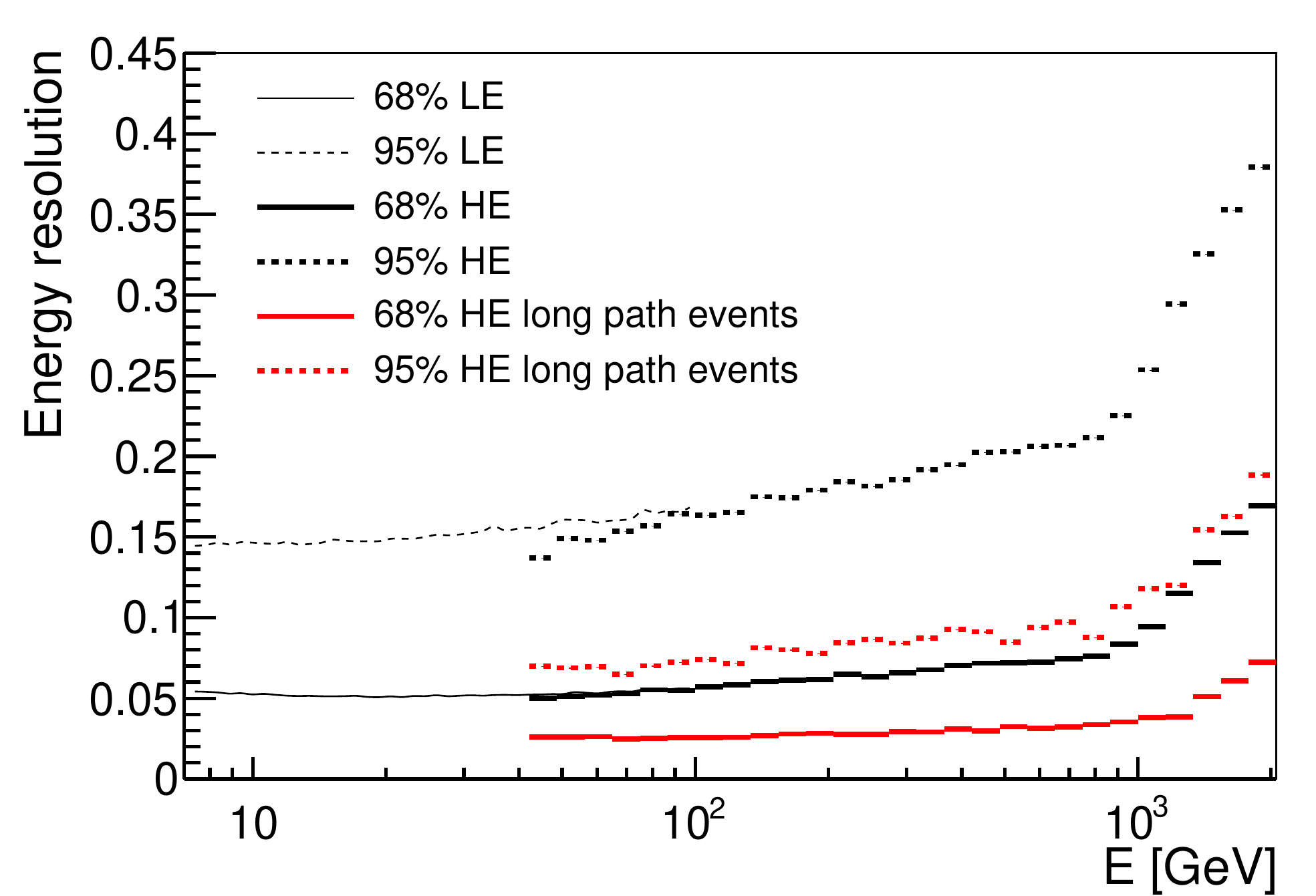}
  \caption{Energy resolution for various selections: all events (black), events with more than 12 radiation lengths in the CAL (red). Solid (dotted) lines correspond to the 68\% (95\%) containment halfwidth. Thin and thick lines correspond to the LE and the HE analysis, respectively.}
  \label{fig:energyresolution}
\end{figure}

Pass~8 improves the energy reconstruction and extends the energy range up to at least 2~TeV. Fig.~\ref{fig:energyresolution} shows that the energy resolution (defined as the half-width of the 68\% containment range) ranges from 4\% at 10~GeV to 8\% at 800~GeV. Above 800~GeV the energy reconstruction is more difficult because of both low shower containment and crystal saturation. As a result the energy resolution increases more rapidly up to 17\% at 2~TeV.

Compared to the previous CRE LAT analysis~\cite{CRE-PRD-LAT}, the energy resolution is significantly improved: at 1~TeV the 68\% and 95\% containment half-widths were 14\% and 34\%, respectively. With Pass 8 they are 10\% and 25\%, while the gain in acceptance is 50\%.

In order to define a subclass of events with a better energy resolution, that is used in section~\ref{sec:results} to test the sensitivity of the spectrum to energy resolution, we select events with a CAL path-length greater than 12~$\mathrm{X}_0$. It corresponds to $\sim 15\%$ of the whole dataset. The average CAL path-length of this long-path-events selection is 13.3~$\mathrm{X}_0$ and the energy resolution is 4\% at 1~TeV and 8\% at 2~TeV.

Because the CAL is only 8.6~$\mathrm{X}_0$ long on-axis, the showers of electrons in the energy range of our analysis are not fully contained. The shower leakage is corrected for by the energy reconstruction. Therefore we have to consider two independent sources of systematic uncertainties for the measured energy. The first one is the uncertainty of the absolute energy scale and the second one is the uncertainty induced by the energy reconstruction. The estimation of these uncertainties is described below.

The geomagnetic cutoff in the CRE spectrum at about 10~GeV provides a spectral feature that allows an absolute calibration of the LAT energy scale. At this energy, the leakage from the CAL is $\sim 20\%$ and the shower maximum lies in the middle of the CAL. As a consequence, the shower profile fit allows a precise energy reconstruction and the systematic uncertainty on the energy reconstruction at 10~GeV is negligible.

A previous measurement of the geomagnetic cutoff was used to check the LAT energy scale based on one year of LAT data~\cite{LAT_absolute_energy_scale}. The same analysis, using almost 7 years of Pass 8 data, is reported in Appendix~B. We find that the ratio of the measured to expected geomagnetic cutoff is $1.033 \pm 0.004  \; (\text{stat}) \pm 0.020 \; (\text{syst})$.
As a result, we decrease the event energies in data by $-3.3\%$ in both the LE and HE analyses and conclude that the systematic uncertainty on the absolute energy scale is 2\%.

\begin{figure}[!htb]
  \includegraphics[width=\linewidth]{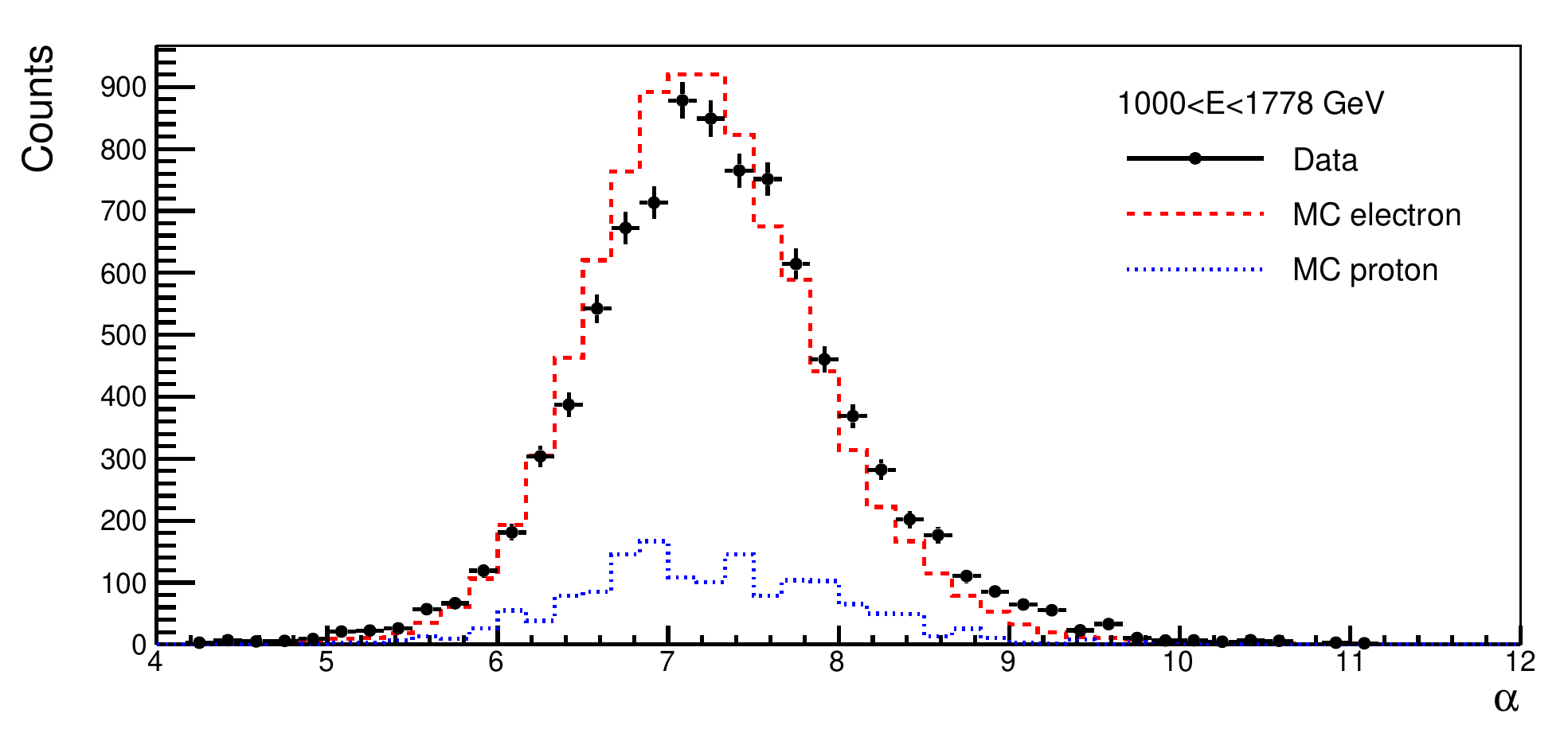}
  \includegraphics[width=\linewidth]{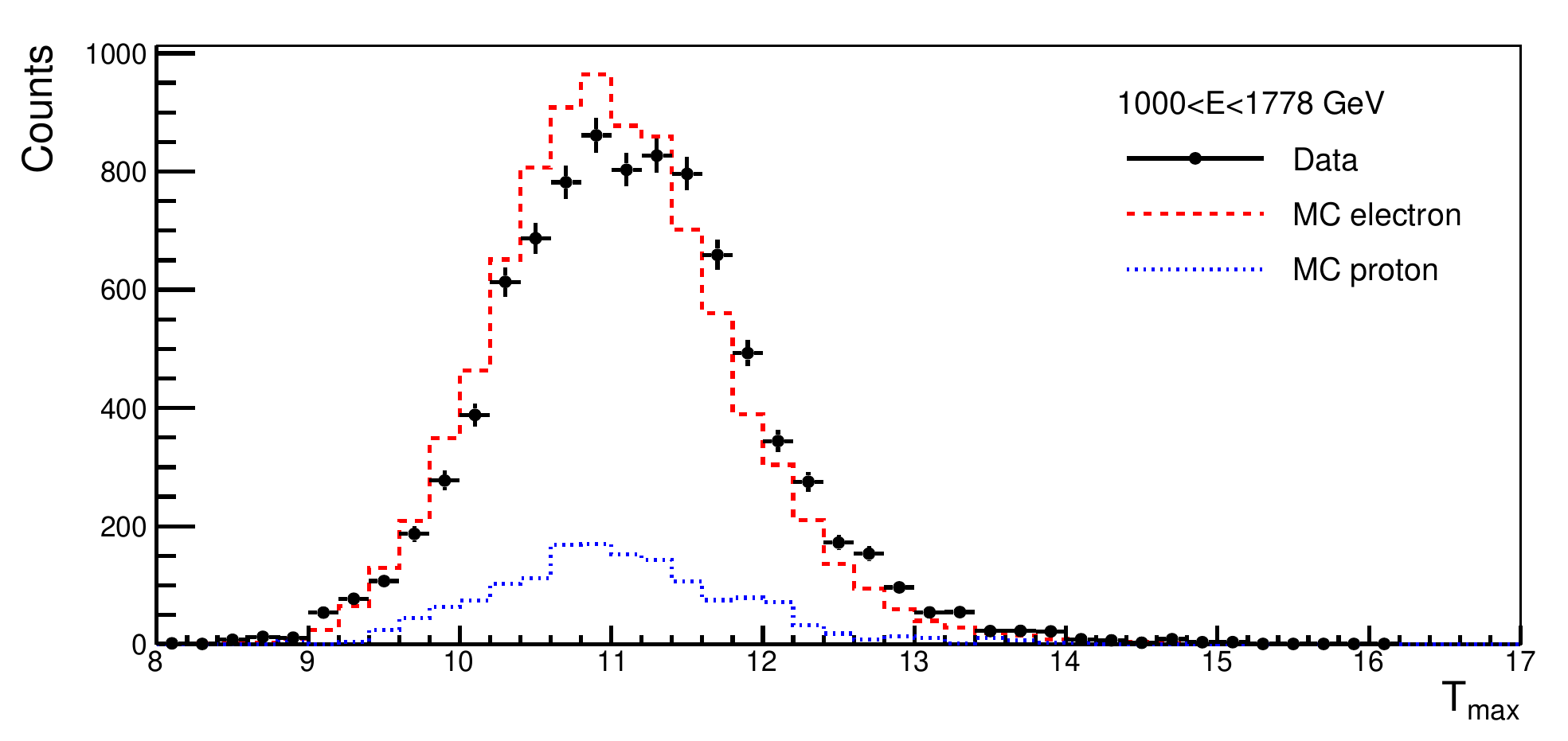}
  \caption{The shower profile parameters $\alpha$ (top) and $T_\mathrm{max}$ (bottom) for events between 1 and 1.78~TeV. The contribution from residual background (blue) has been subtracted from the data distributions. The red histograms correspond to the electron simulation.}
  \label{fig:showvar_CalNewCfpAlpha_CalNewCfpTmax}
\end{figure}

Above 10~GeV the shower leakage increases linearly with $\log_{10}E$ and the shower maximum gets closer to the end of the CAL. Therefore we expect that a potential systematic bias on the energy reconstruction would increase with the energy as well. For each event, the rear leakage corresponds to the extrapolation of the shower profile beyond the total amount of radiation lengths seen by the shower. The precision of this extrapolation depends on the precision we have on the parameters of the fit that drive the shower shape. In order to assess this precision, we compare the distributions of $\alpha$ and $T_\mathrm{max}$ in data and in the simulation, as shown in Fig.~\ref{fig:showvar_CalNewCfpAlpha_CalNewCfpTmax} for events between 1 and 1.78~TeV. We find that the data/MC differences as a function of energy for $\alpha$ and $T_\mathrm{max}$ are respectively within $\pm\delta_\alpha(E)$ and $\pm \delta_T(E)$, with $\delta_\alpha(E) = 0.05\log_{10}(E/\text{[10 GeV]})$ and $\delta_T(E) = 0.1\log_{10}(E/[10 GeV])$.

We estimate the uncertainty due to shower leakage by varying $\alpha$ and $T_\mathrm{max}$ within $\pm\delta_\alpha(E)$ and $\pm\delta_T(E)$, renormalizing the modified profile so that it matches the last layer energy and computing the maximum leakage variation. We find that it varies linearly with $\log_{10}E$ and that the resulting variation of the total reconstructed energy varies linearly with $\log_{10}E$ as $\delta E_\mathrm{rec}(E) = 0.025\log_{10}(E/[10 GeV])$, which is 5\% at 1~TeV. We have checked that saturation does not contribute significantly to this systematic uncertainty, as explained in Appendix~A.

\section{Systematic uncertainties}
\label{sec:systematics}

For the HE analysis, we consider four sources of systematic uncertainty. The first three (acceptance, contamination, IVC) relate to the event selection, while the last one is the uncertainty of the energy measurement.

The uncertainty on the acceptance is estimated in each energy bin by measuring the sensitivity of the measured CRE flux to the choice of the cut on $P_\text{CRE}$ by varying $P_\text{cut}$ in a range corresponding to a variation in efficiency of $\pm20\%$ around the nominal efficiency (under the requirement that $P_\text{cut}<P_\text{max}$). The flux variation, which we attribute to a remaining data/MC disagreement, is found to be less than 2\% up to $\sim 500$~GeV, increasing to 6\% at 2~TeV.

The number of residual background events is estimated by fitting simulated background templates to the data. In order to take into account the uncertainty of the Geant4 prediction of the fraction of protons mimicking electromagnetic showers, we assume a 20\% uncertainty on the number of background events after the selection cut~\cite{Yarba:2012ih,Dotti:2013gya,Bilki:2014bga}. Due to the small residual background contamination, this uncertainty leads to a change in the number of signal events of less than 2\% up to 1~TeV, increasing to 7\% at 2~TeV.

\begin{figure}[!htb]
  \includegraphics[width=\linewidth]{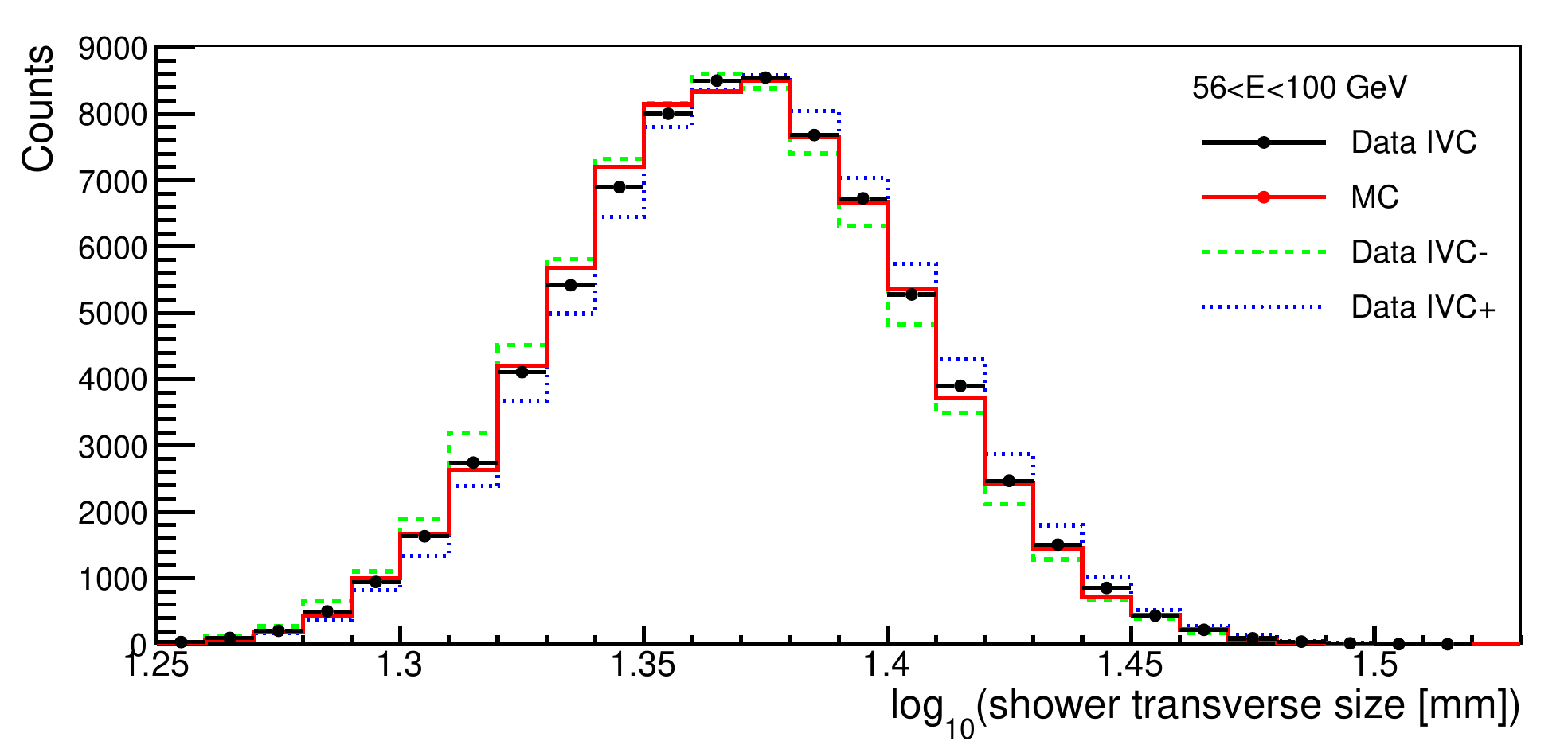}
  \includegraphics[width=\linewidth]{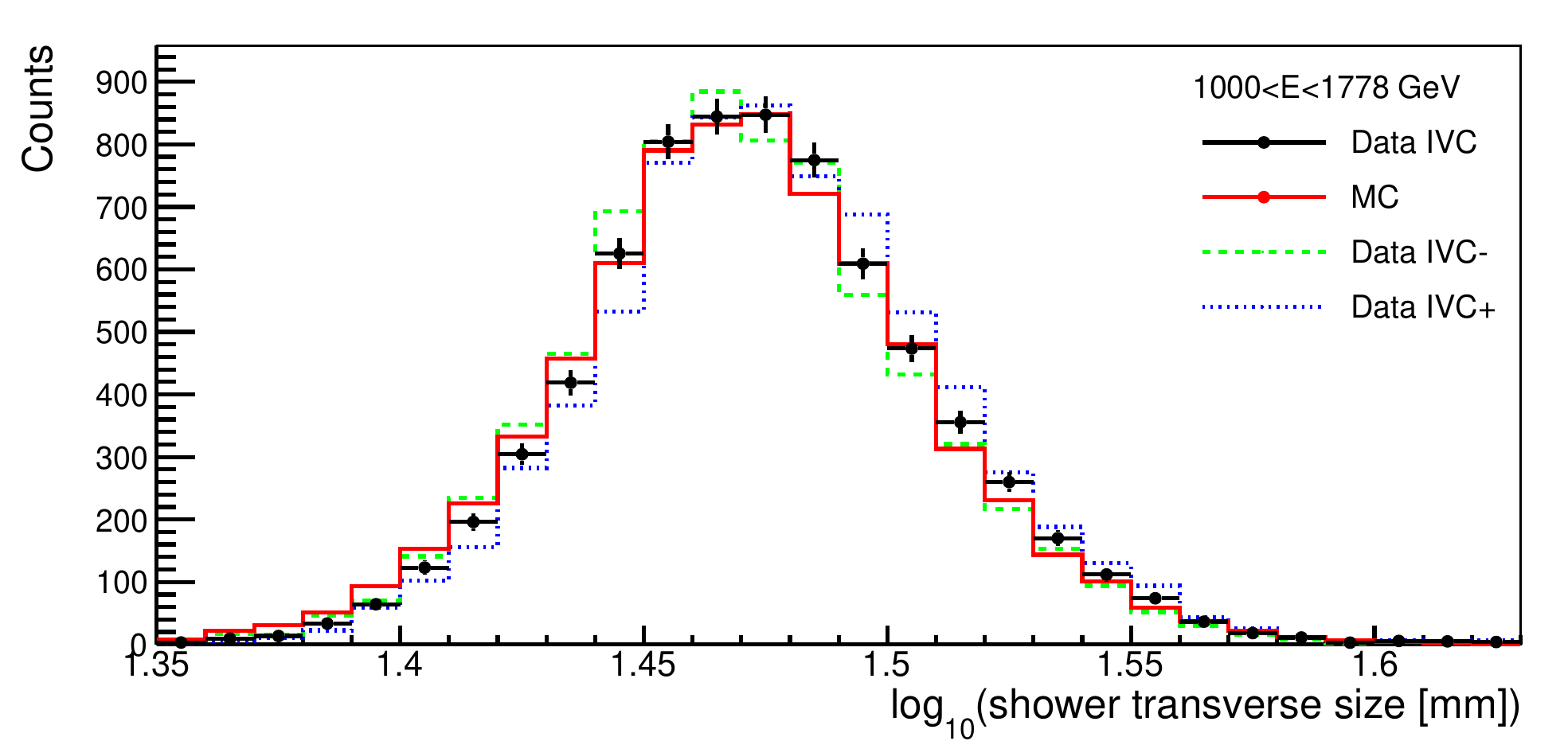}
  \caption{The logarithm of the shower transverse size after the nominal IVC correction (\emph{black}) and after moving the correction factors by plus or minus the maximum data/MC residual differences (\emph{blue} and \emph{green}) for events between 56 and 100~GeV (top) and between 1 and 1.78~TeV (bottom). The contribution from residual background has been subtracted from the data distributions. The red histograms correspond to the electron simulation.}
  \label{fig:syst_cordat}
\end{figure}

For each input variable to the BDT, the IVC corrections are derived from the difference between the peak position of the data and MC distributions. After IVC corrections, there are still some small residual differences. We derive two alternative sets of IVC corrections in which each correction is displaced by plus or minus the maximum of the residual differences at any energy and inclination angle. Fig.~\ref{fig:syst_cordat} compares the distributions of the transverse size of the shower obtained with these two sets of corrections to the nominal one and the prediction of the simulation. The variation of the number of signal events compared to the nominal IVC corrections increases from 2\% at 42~GeV to 10\% at 1~TeV, reaching 14\% at 2~TeV.

Regarding the systematic uncertainty on the energy measurement, there are two independent sources of uncertainty as presented in section~\ref{sec:energymeasurement}. The first one is the systematic uncertainty on the absolute energy scale, which does not depend on energy and is 2\%. The second one is the systematic uncertainty on the energy reconstruction and varies linearly with $\log_{10}E$ from 0\% at 10~GeV to 5\% at 1~TeV.

In order to account for the energy dependent part of the energy measurement systematic uncertainty, we change the energies of all flight-data events according to some conservative scenarios that depend on the spectral hypothesis we test and repeat the whole analysis. Compared to this uncertainty, the constant 2\% uncertainty on the absolute energy scale is subdominant and is not considered when fitting the CRE spectrum.

When fitting the CRE spectrum in the HE analysis energy range, we add in quadrature the acceptance uncertainty to the statistical uncertainty and we treat the sum of the contamination and the IVC corrections uncertainties as nuisance parameters, as described in Appendix~C. When displaying the spectrum, the statistical and systematic uncertainties (except the one on the energy measurement) are added in quadrature.

For the LE analysis, we consider the acceptance and contamination systematic uncertainties, as well as the changes induced in the spectrum by changing the McIlwain~L selection. The sum of these uncertainties is $\leq~4\%$ and is added in quadrature to the statistical uncertainty.

\section{Results and discussion}
\label{sec:results}

\begin{figure}[ht]
  \includegraphics[width=0.8\linewidth]{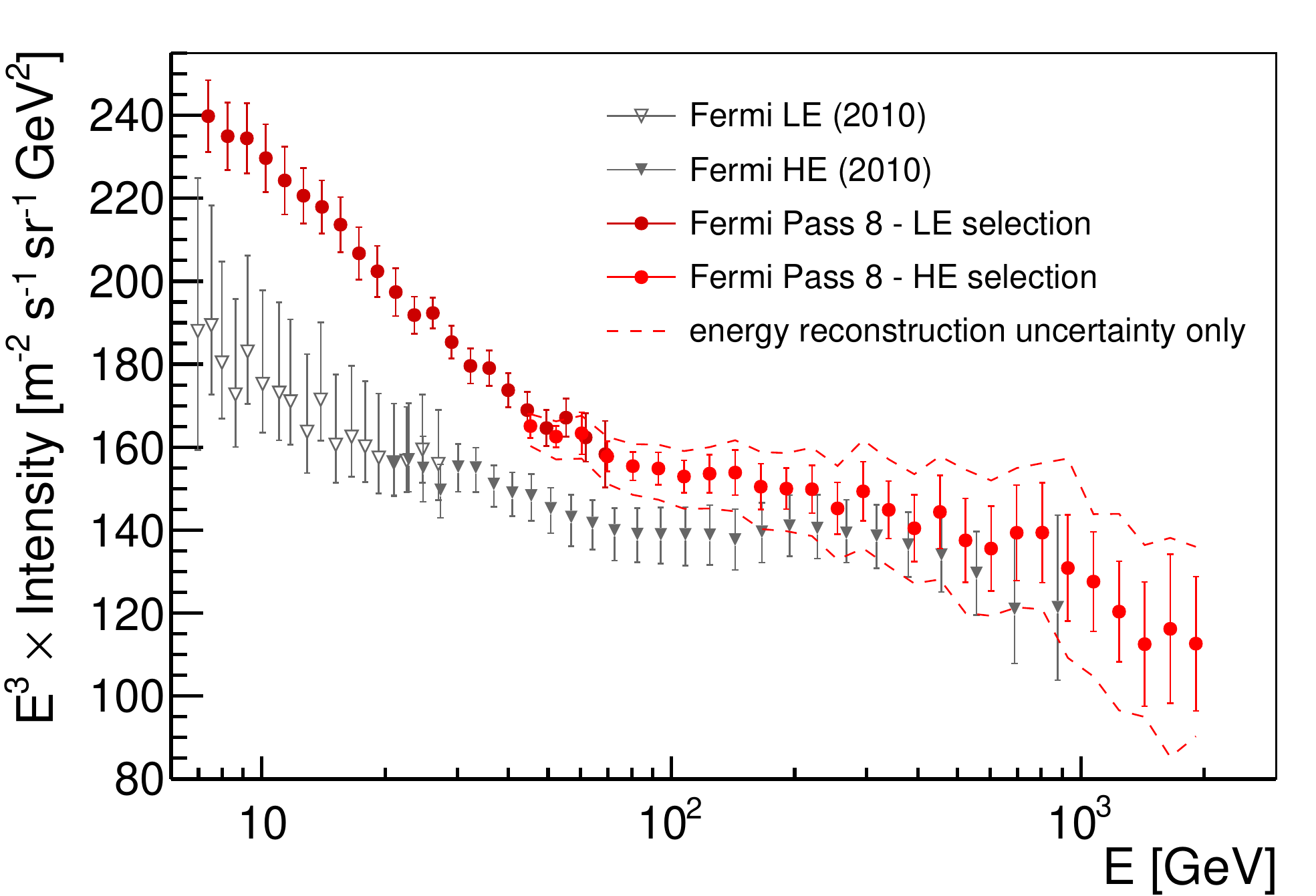}
  \caption{CRE spectrum between 7~GeV and 2~TeV measured by the LAT and the previous LAT measurement~\cite{CRE-PRD-LAT}. All error bars represent the quadratic sum of statistical and systematic uncertainties (except the one on the energy measurement). The LAT flux is multiplied by the cube of the representative energy in each bin, computed following Eq.~(6) of~\cite{Lafferty:1994cj} with an $E^{-3}$ spectrum. The area between the dashed lines corresponds to the uncertainty band due to the LAT energy reconstruction uncertainty only. The 2\% systematic uncertainty on the energy scale is not indicated.}
  \label{fig:latoldnewspectrum}
\end{figure}

For both the LE and HE analyses, we fit the CRE count spectrum by forward folding the input flux using the Detector Response Matrix (DRM) in order to take into account the detector energy resolution. The results are shown in Fig.~\ref{fig:latoldnewspectrum}~and~\ref{fig:finalspectrum}. Tables with fluxes as well as event numbers can be found in Appendix~D. The bin-by-bin fluxes are obtained by performing a fit with the DRM in each bin separately with a single power law with a fixed $3.1$ spectral index. We note that the LE and HE spectra match very well over the overlapping range $42<E<70$~GeV.

The dashed lines illustrate the systematic uncertainty on the energy reconstruction and correspond to the central values of the LAT flux for two scenarios in which the energy is changed by a factor that varies linearly in $\log_{10}E$ between $0\%$ at 10~GeV and $\pm 5\%$ at 1~TeV. We emphasize that these lines do not take into account the statistical uncertainty nor the systematic uncertainties unrelated to the energy measurement. 

\begin{figure}[!htb]
  \includegraphics[width=.8\linewidth]{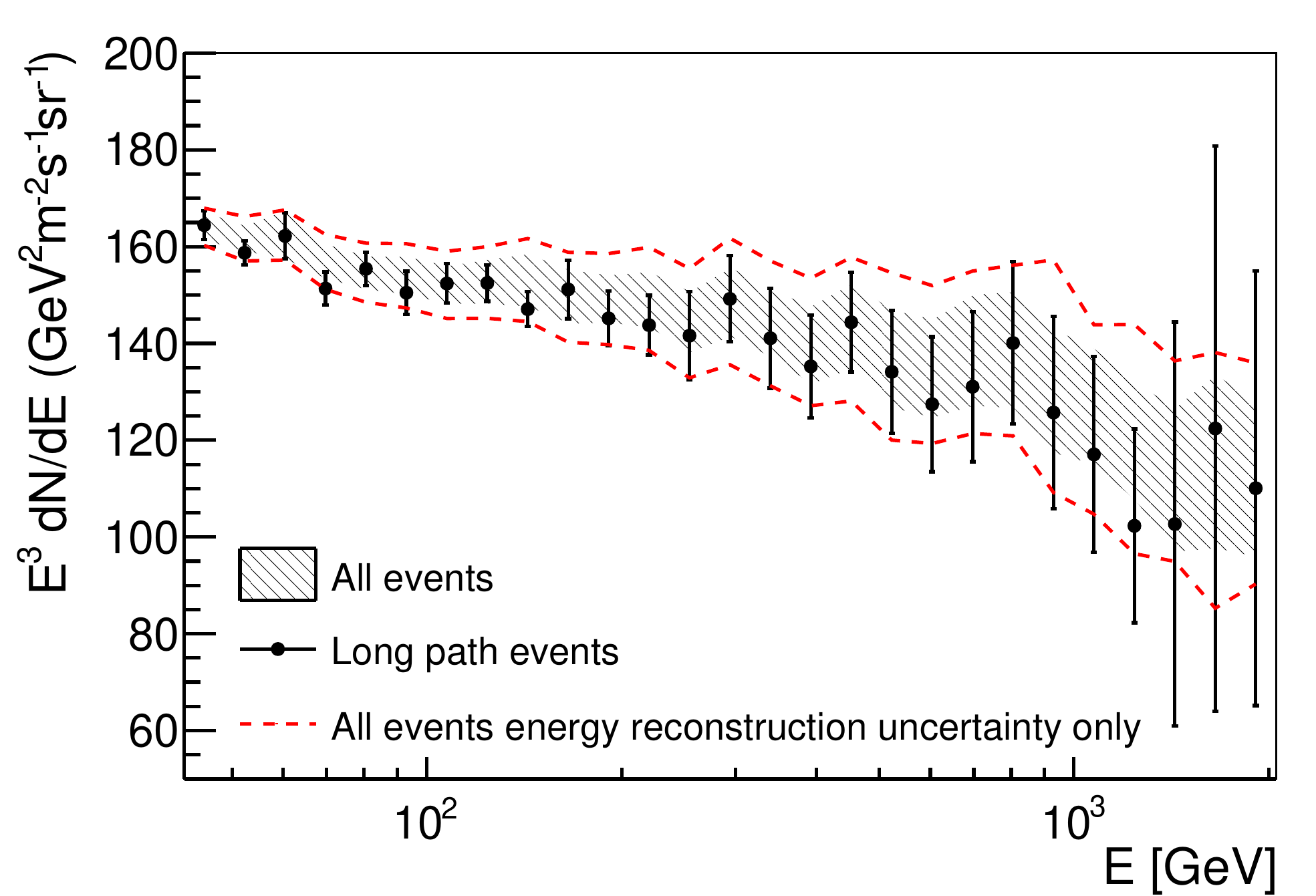}
  \caption{CRE spectrum between 42~GeV and 2~TeV measured with all events (grey band) and with long-path events (black points). In both cases, the statistical and systematic uncertainties (except for the energy measurement) are added in quadrature. The area between the dashed lines corresponds to the uncertainty band due to the LAT energy reconstruction uncertainty only of the all-event selection.}
  \label{fig:spectrumonoffaxis}
\end{figure}

We also derive the HE spectrum using the subclass of events with a path length in the CAL greater than 12~$\mathrm{X}_0$ introduced in section~\ref{sec:energymeasurement}. This spectrum is systematically lower than but compatible with the all-events spectrum, as can be seen in Fig.~\ref{fig:spectrumonoffaxis}. Although the energy resolution for the long-path selection is much better, the systematic uncertainties (except the one on the energy measurement) are similar to the ones of the all-events spectrum up to 200~GeV and larger above. Regarding the systematic uncertainty on the energy measurement, the long-path selection spectrum is halfway between the nominal spectrum and the spectrum corresponding to an energy correction of -5\% at 1~TeV. It is compatible with the systematic uncertainty on the energy measurement of the long-path selection which is 2.5\% at 1~TeV. We conclude that the long-path selection does not allow a more precise measurement of the CRE spectrum but it tends to favor a scenario in which the energy correction is negative rather than positive.

Below 100~GeV, the new LAT measurement differs from the previous one by 10--30\%, as can be seen in Fig.~\ref{fig:latoldnewspectrum}. A large part of this difference below 30~GeV is due to the lack of correction in the previous analysis for the loss of CREs above the geomagnetic energy cutoff. After applying this correction, the remaining difference is 10--15\% and is due to imperfections in the simulation that was used in the previous analysis (remnants of electronic signals from out-of-time particles were not simulated~\cite{FermiInstrument2012}).

The CRE spectrum between 7 and 42~GeV is well fitted by a power law with a spectral index $3.21 \pm 0.02$. The low $\chi^2$ (2.25 for 15 degrees of freedom) means that the systematic uncertainties are too large to detect the deviation from a power law due to the magnetic field of the heliosphere. This is strengthened by the fact that fitting between 15 and 42~GeV changes the spectral index by only 0.005. We therefore do not take into account the heliospheric effects in the following fits.

As can be seen in Fig.~\ref{fig:finalspectrum}, when not taking into account the uncertainty on the energy reconstruction, the LAT CRE spectrum is above the AMS-02 one for energies larger than $\sim 70$~GeV and suggests the presence of a break in the spectrum. Fitting the spectrum between 7~GeV and 2~TeV with a single power law yields $\chi^2 = 64.6$ for 36 degrees of freedom, corresponding to a probability of $2.4 \times 10^{-3}$. As expected, a broken power-law fit yields a much lower $\chi^2=19.2$ for 34 degrees of freedom. The break energy is $53 \pm 8$~GeV and the spectral indices below and above the break are $3.21 \pm 0.02$ and $3.07 \pm 0.02$, respectively.

\begin{figure}[ht]
  \includegraphics[width=1.0\linewidth]{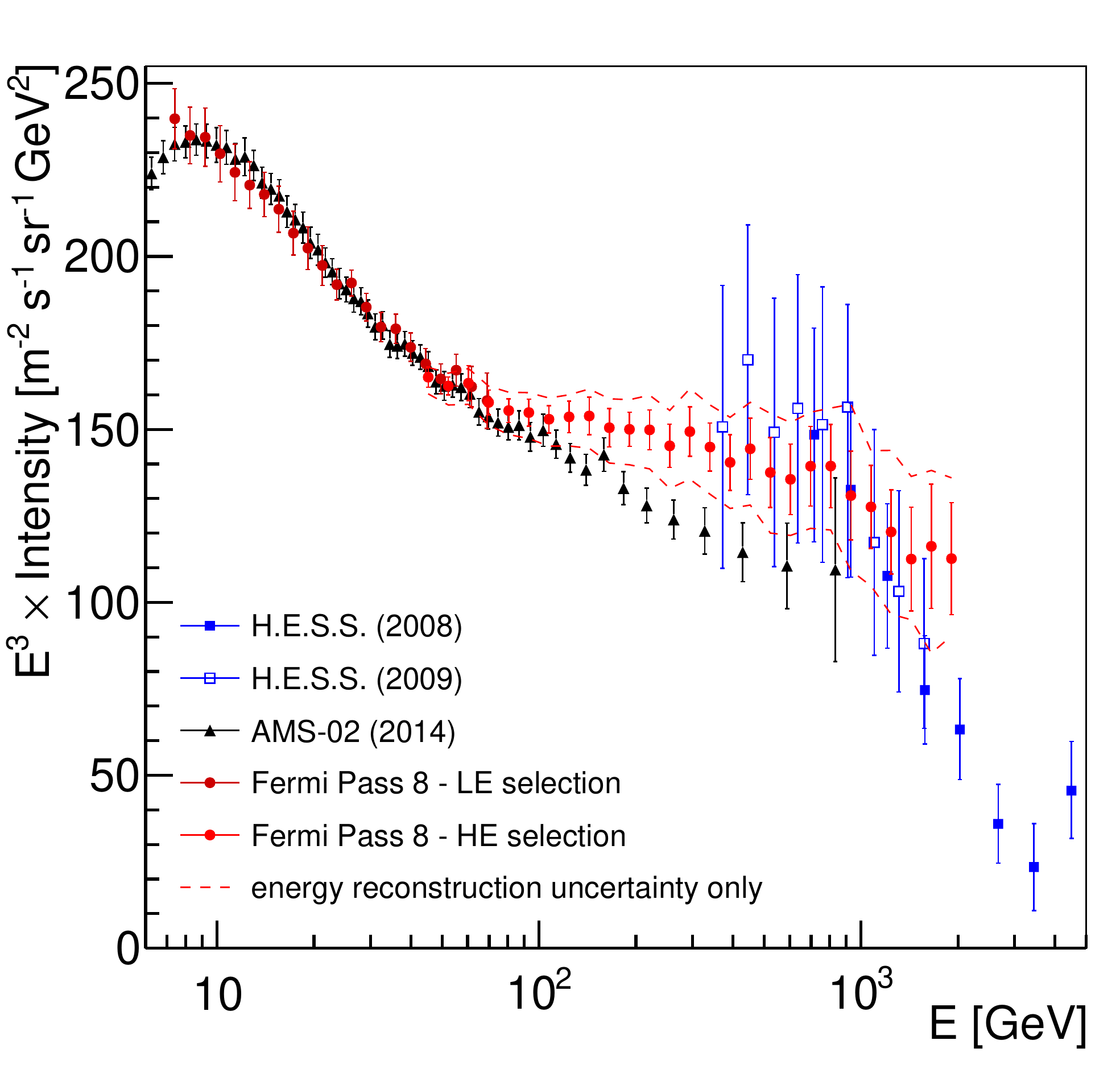}
  \caption{CRE spectrum between 7~GeV and 2~TeV measured by the LAT along with other recent measurements by AMS-02~\cite{AMS_PRLnov2014} and H.E.S.S.~\cite{HESS2008,HESS2009}. All error bars represent the quadratic sum of statistical and systematic uncertainties (except the one on the energy measurement). The LAT flux is multiplied by the cube of the representative energy in each bin, computed following Eq.~(6) of~\cite{Lafferty:1994cj} with an $E^{-3}$ spectrum. The area between the dashed lines corresponds to the uncertainty band due to the LAT energy measurement uncertainty only. The 2\% systematic uncertainty on the energy scale is not indicated.}
  \label{fig:finalspectrum}
\end{figure}

In order to estimate the influence of the energy measurement systematic uncertainty on the detection of a break, we consider the event-energy rescaling scenario that would be responsible for such a break, among the scenarios allowed by the systematic uncertainty on the energy reconstruction. In this scenario the energy is unchanged up to 50~GeV and then decreases linearly with $\log_{10}E$ to $-5\%$ at 1~TeV. The single power-law fit then yields a $\chi^2=49.9$ for 36 degrees of freedom, corresponding to a 6\% probability.

This relatively low probability suggests that the broken power-law hypothesis would be preferred. The broken power-law fit, performed with the same scenario, yields indeed a lower $\chi^2=28.9$ for 34 degrees of freedom, a break energy of $47 \pm 6$~GeV and spectral indices below and above the break of $3.21 \pm 0.02$ and $3.11 \pm 0.02$, respectively. The $\chi^2$ difference is 21 for two less degrees of freedom. The broken power-law hypothesis is thus preferred at the $4\sigma$ level. We note that in all the fits, some of the systematic uncertainties are treated using the nuisance parameter approach. Therefore the $\chi^2$ and corresponding probabilities depend slightly on the number of nuisance parameters but the level of significance of the break does not change.

AMS-02 estimated the lower limit above which the flux is described by a single power law and found 30.2~GeV~\cite{AMS_PRLnov2014}, reporting a spectral index above this energy of $3.170 \pm 0.008 \; (\mathrm{stat+syst}) \pm 0.008 \; (\text{energy measurement})$. Performing a single power-law fit of the LAT spectrum above the same energy with the energy modification scenario introduced above, we find a spectral index of $3.125 \pm 0.020$, with a $\chi^2$ of 28.9 for 22 degrees of freedom, corresponding to a 14\% probability. Comparing this result to the lower AMS-02 value allowed by the systematic uncertainty on the energy scale, we find that the difference between the Fermi and AMS-02 spectral indices above 30.2~GeV is $0.037 \pm 0.022$. The difference is at the level of $1.7\sigma$. This could indicate that systematic uncertainties on the energy measurement in one or both of the results are slightly larger than estimated.

At higher energies, H.E.S.S. reported in~\cite{HESS2008} that, leaving its energy scale factor free, the H.E.S.S. data above 600~GeV combined with earlier data were well reproduced by an exponentially cutoff power law with an index of $3.05 \pm 0.02$ and a cutoff at $2.1 \pm 0.3$~TeV. The LAT CRE spectrum above 50~GeV, as indicated by the previous broken power-law fits, is compatible with a single power law with a spectral index of $3.07 \pm 0.02 \; (\text{stat+syst}) \pm 0.04 \; (\text{energy measurement})$. Fitting the count spectrum above 50~GeV with an exponentially cutoff power law $E^{-\gamma} e^{-E/E_c}$ does not yield statistically significant evidence for a cutoff (a $\chi^2$ difference of 1 for 1 less degree of freedom) and we exclude $E_c<2.1$~TeV at 95\% CL. Assuming a scenario in which the energy is changed by 0\% at 50~GeV to $-5\%$ at 1~TeV, we exclude $E_c<1.8$~TeV at 95\% CL.

Regarding the agreement between the H.E.S.S. and LAT spectra, we note that, as can be seen in Fig.~\ref{fig:finalspectrum}, the energy measurement scenario corresponding to the lower dashed line yields a LAT CRE spectrum that connects to H.E.S.S. data around 1~TeV. With this scenario, the LAT spectral index is $\sim 3.11$, relatively steeper than $3.05$, as reported by H.E.S.S. As a result, a LAT and H.E.S.S. combined fit would lead to a cutoff larger than $\sim 2.1$~TeV, well above the LAT lower limit of $1.8$~TeV.

The precision of the LAT measurement is limited primarily by the energy dependent systematic uncertainty on the energy measurement, due to the low containment of CRE induced showers in the LAT calorimeter that worsens with energy. A possible way to mitigate this issue would be to use CRE events with an incidence angle greater than 60~degrees. The showers of these events are much more contained in the LAT instrument and the systematic uncertainty on the energy reconstruction would thus be reduced. Unfortunately, the drawback of this approach is that the track information of such events is scarce or inaccurate, which strongly hampers background rejection. We have started to investigate this approach but assessing the improvement that it can lead to is beyond the scope of this paper.

If this approach proves to be successful, its results, along with updated measurements of AMS-02 and H.E.S.S. and the first results of DAMPE~\cite{DAMPE} and CALET~\cite{CALET}, would certainly help in detecting and characterizing precisely the features of the CRE spectrum between 10~GeV and several TeVs.

\section*{acknowledgements}

The \textit{Fermi}-LAT Collaboration acknowledges support for LAT development, operation and data analysis from NASA and DOE (United States), CEA/Irfu and IN2P3/CNRS (France), ASI and INFN (Italy), MEXT, KEK, and JAXA (Japan), and the K.A.~Wallenberg Foundation, the Swedish Research Council and the National Space Board (Sweden). Science analysis support in the operations phase from INAF (Italy) and CNES (France) is also gratefully acknowledged.

We would like to thank the INFN GRID Data Centers of Pisa, Trieste and CNAF-Bologna, the DOE SLAC National Accelerator Laboratory Computing Division and the CNRS/IN2P3 Computing Center (CC-IN2P3 - Lyon/Villeurbanne) in partnership with CEA/DSM/Irfu for their strong support in performing the massive simulations necessary for this work.

W. Mitthumsiri is partially supported by the Thailand Research Fund (Grants TRG5880173 and RTA5980003).

\section*{Appendix A: Energy reconstruction}

The LAT energy reconstruction above $\sim 5$~GeV is performed by fitting the shower profile, using the 8 CAL layer energies. We use the following representation of the longitudinal shower profile~\cite{Longo:1975wb}
\begin{equation}
\frac{dE(t)}{dt} = E \times \mathcal{P}(\alpha,\beta,t) = E \times \frac{(\beta t)^{\alpha-1} \beta e^{-\beta t}} {\Gamma(\alpha)}
\label{eq:showerprofile}
\end{equation}
where $t$ is the longitudinal shower depth in units of radiation length, $\alpha$ the shape parameter and $\beta$ the scaling parameter. The profile shape $\mathcal{P}$ is such that $\int \mathcal{P}(\alpha,\beta,t)dt = 1$ and its maximum for given values of $\alpha$ and $\beta$ is reached at $T_\mathrm{max} = (\alpha -1)/\beta$. The profile fit has 3 parameters (the energy and two shape parameters) and 5 degrees of freedom.

We use a model of the longitudinal profile (mean and standard deviation of the shape parameters) and a model of the average radial profile of electromagnetic showers in CsI. Both models describe the variation of the longitudinal and radial profiles with energy and were derived using dedicated Geant4 simulations from 1~GeV to 3~TeV.

The profile fit fully takes into account the geometry of the LAT calorimeter (especially the gaps between modules) in order to predict the energy deposited in the layers and crystals for any given shower profile. The energy deposition prediction is performed on an event-by-event basis, by going forward along the event axis (measured with the tracker) in steps of 1.85~mm ($0.1 \; \mathrm{X}_0$). At each step, we compute the fraction of energy deposited in each crystal, taking into account the shower longitudinal and radial profiles and the calorimeter geometry.

The fraction of energy deposited in the crystals as a function of distance along the event axis is translated into a fraction of deposited energy as a function of radiation length, which is used in the profile fit to compute the layer energies that are compared to the measured ones. It is to be noted that the longitudinal profile is free to fluctuate in the fit according to the model of the longitudinal profile derived with Geant4. This is done by adding to the $\chi^2$ a gaussian prior for each of the shape parameters, with mean and standard deviation as given by the model of the longitudinal profile. More information can be found in~\cite{EnergyReconProceedings}.

Pass 8 introduces several improvements to the energy reconstruction:
\begin{itemize}
\item the upper end of the energy range over which the shower longitudinal and radial models have been computed has been increased from 150~GeV to 3~TeV;
\item the widening of the radial profile in the gaps between modules has been modeled;
\item in the Pass 6/7 version of the profile fit, layers with at least one saturated crystal were discarded. In Pass 8, only the saturated crystals are discarded: for each layer, the energy that is considered in the fit is the sum of the energy of the non-saturated crystals.
\end{itemize}

The last point must be taken into account when estimating the systematic uncertainty on the energy reconstruction presented in section~\ref{sec:energymeasurement}. The energy deposited in the saturated crystals is missed and this additional leakage amounts on average to 10\% at 1~TeV, increasing to 25\% at 2~TeV. Saturation occurs for crystals in the core of the shower. Therefore, the predicted energy for the layers with saturated crystals depends on the radial profile model that we use.

In order to quantify the dependence of the energy reconstrution on the radial profile model, we scale it by $\pm20\%$. As expected, scaling the radial profile changes the $\chi^2$ of the fit but we find that this variation occurs only above $\sim 800$~GeV, that is to say when crystal saturation is important. Below $\sim 800$~GeV the $\chi^2$ is unchanged. The comparison of the $\log_{10}{\chi^2}$ between data and MC shows that the data/MC difference increases linearly with $\log_{10}E$, from 0 at 10~GeV to 0.13 at 1~TeV, without any sharp variation around 800~GeV. Quantitatively, the variation of this data/MC difference between 800~GeV and 1~TeV is less than 0.05, which would correspond to a 5\% scaling of the radial profile. We therefore conclude that the radial profile model we use is correct within 5\%. When scaling the radial profile within 5\%, we find that the reconstructed energy variation is smaller than $0.1\delta E_\mathrm{rec}(E)$ and conclude that saturation does not significantly contribute to the systematic uncertainty on the energy reconstruction.

\section*{Appendix B: Measurement of the absolute energy scale at $\sim$10~GeV}

The geomagnetic cutoff in the CRE spectrum provides a spectral feature that allows an absolute calibration of the LAT energy scale. A previous measurement of the geomagnetic cutoff was used to calibrate the LAT energy scale based on one year of LAT data~\cite{LAT_absolute_energy_scale}. We performed the same analysis, using almost 7 years of Pass 8 data, in 6 McIlwain~L intervals. We used the LE CRE estimator with a cut ensuring a constant 2\% residual background contamination. The $P_\text{CRE}$ cut efficiency for electrons is $\sim 30\%$ below 5~GeV, rising to 90\% at 10~GeV. In order to check the sensitivity of the measurements to the selection, we also used a selection ensuring a constant 5\% residual background contamination, which corresponds to an efficiency of 70\% below 5~GeV, rising to 95\% at 10~GeV.

\begin{figure}[!htb]
  \includegraphics[width=\linewidth]{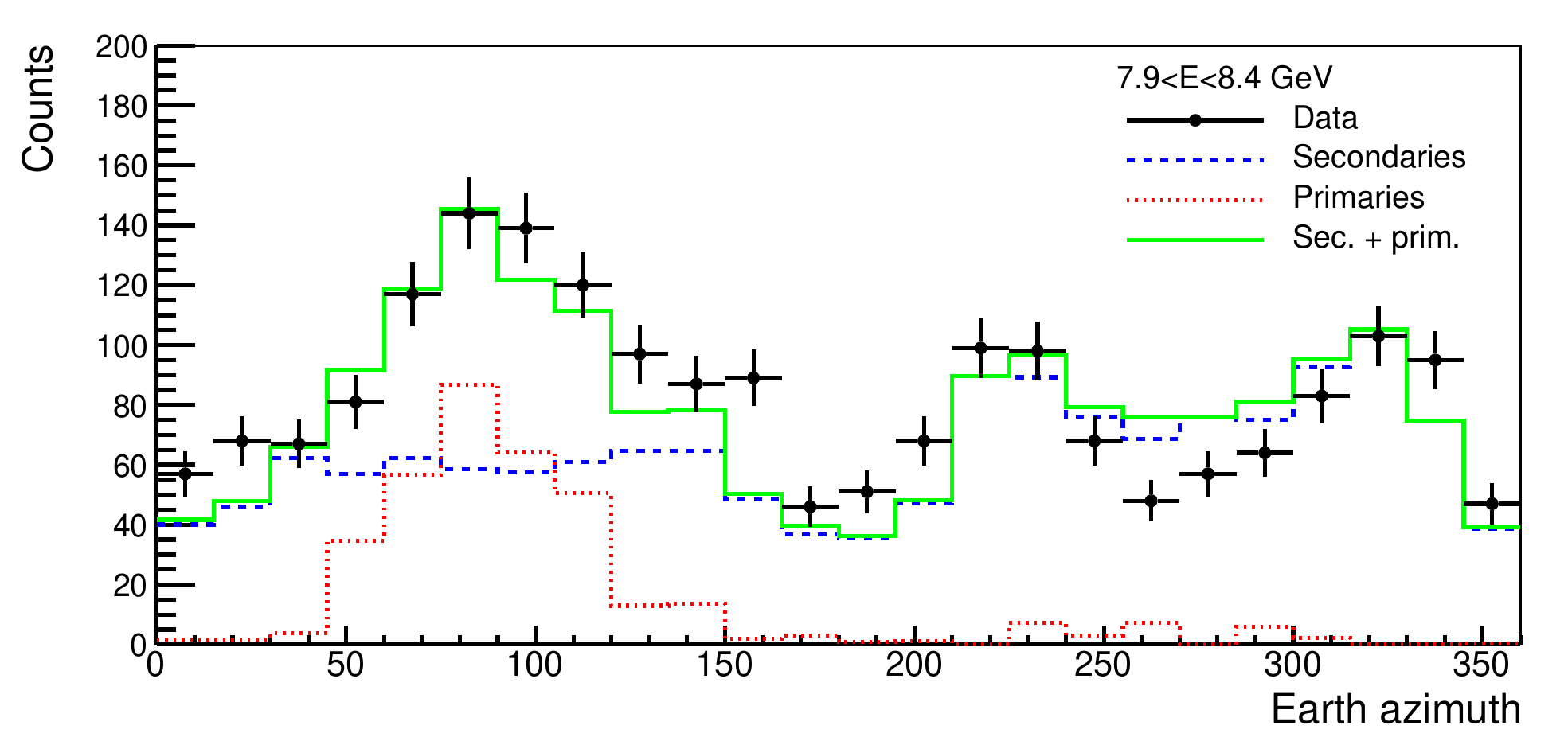}
  \includegraphics[width=\linewidth]{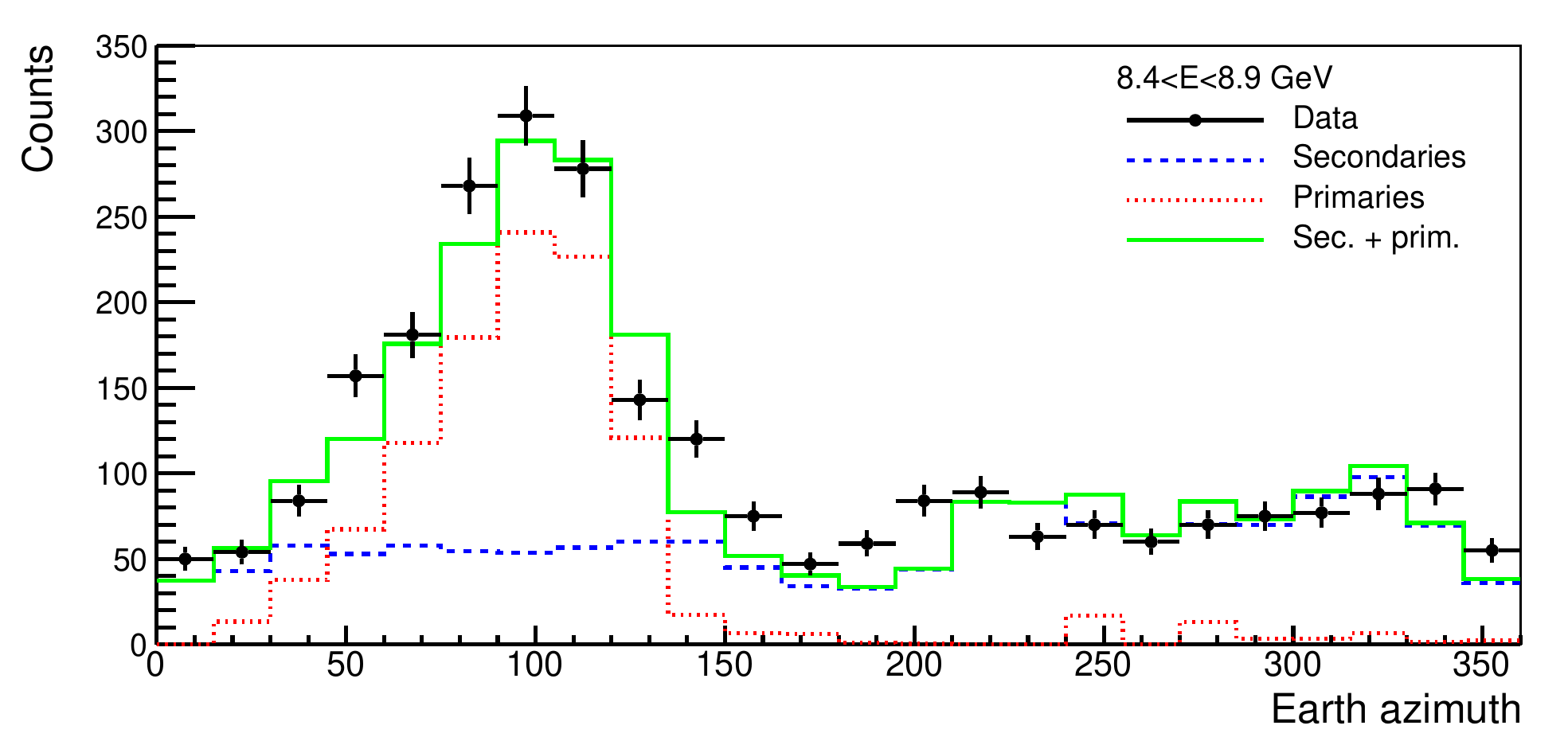}
  \caption{Examples of the template fit of the azimuthal distribution for McIlwain~L in [1.0, 1.1]: events between 7.9 and 8.4~GeV (top) and between 8.4 and 8.9~GeV (bottom). The black, red, blue and green histograms correspond to data, the secondary template, the primary template, and the sum of the secondary and primary templates. North and East correspond to 0 and $90^{\mathrm{o}}$, respectively.}
  \label{fig:absoluteenergyscale_azimtemplatefit}
\end{figure}

After background subtraction, we are left with primary and secondary electrons. In order to measure the fraction of secondaries, we fit the CRE azimuthal distribution with the sum of two templates: for primary CREs, we use the one predicted by \emph{tracer} and, for secondary electrons, we use the one observed in data well below the rigidity cutoff (when the primary fraction is lower than 0.5\%). Fig.~\ref{fig:absoluteenergyscale_azimtemplatefit} shows two examples of such fits for McIlwain~L in [1.0, 1.1]. They correspond to two adjacent energy bins below the geomagnetic cutoff, in which the primary fraction increases from 20\% to 45\%.

\begin{figure}[!htb]
  \includegraphics[width=.8\linewidth]{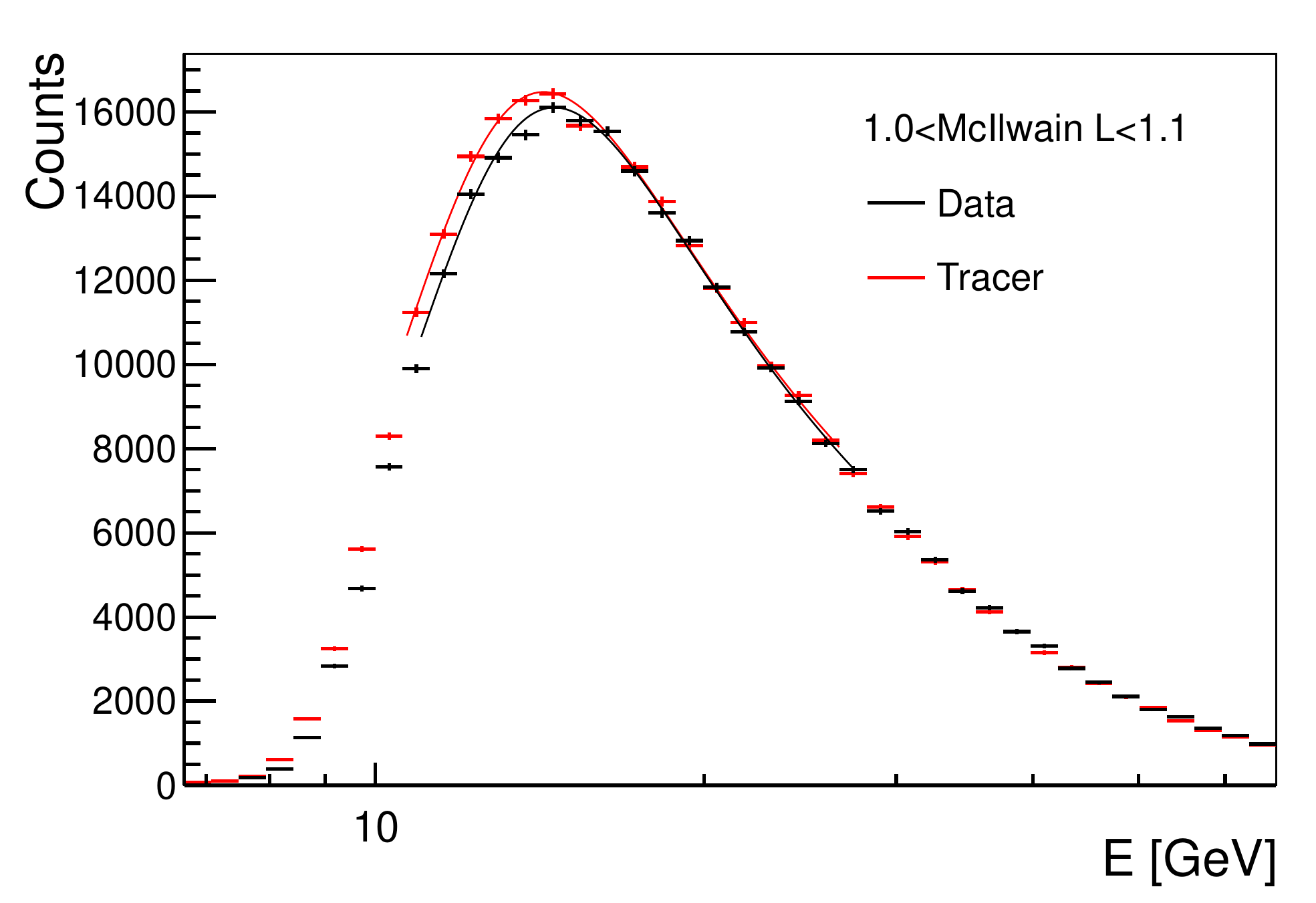}
  \caption{Measured count spectrum (black) of primary CREs after removal of secondary electrons for the McIlwain~L interval [1.0, 1.1], and the count spectrum predicted by \emph{tracer}. The latter is normalized so that its integral above 20~GeV is the same as in the data.}
  \label{fig:absoluteenergyscale_countspectrum}
\end{figure}

For each McIlwain L interval, we fit for the contribution of secondaries as a function of energy and subtract it to obtain the count spectrum of primary CRE. We fit the count spectrum with $E^{{-\gamma}_{p}}/ (1+(E/E_{c})^{-\alpha})$, where $E_{c}$ is the cutoff and ${\gamma}_{p}$ is the spectral index above $E_{c}$. We modify the \emph{tracer} input spectrum according to solar modulation, using the force field approximation with a solar modulation parameter $\phi$ set to an average value for the 2008--2015 period of $500$~MV. Fig.~\ref{fig:absoluteenergyscale_countspectrum} shows the result of the fit for the McIlwain~L interval [1.0, 1.1]. We then compare the values of $E_{c}$ obtained in the data and predicted by \emph{tracer} to check the LAT absolute energy scale.

\begin{figure}[!htb]
  \includegraphics[width=.8\linewidth]{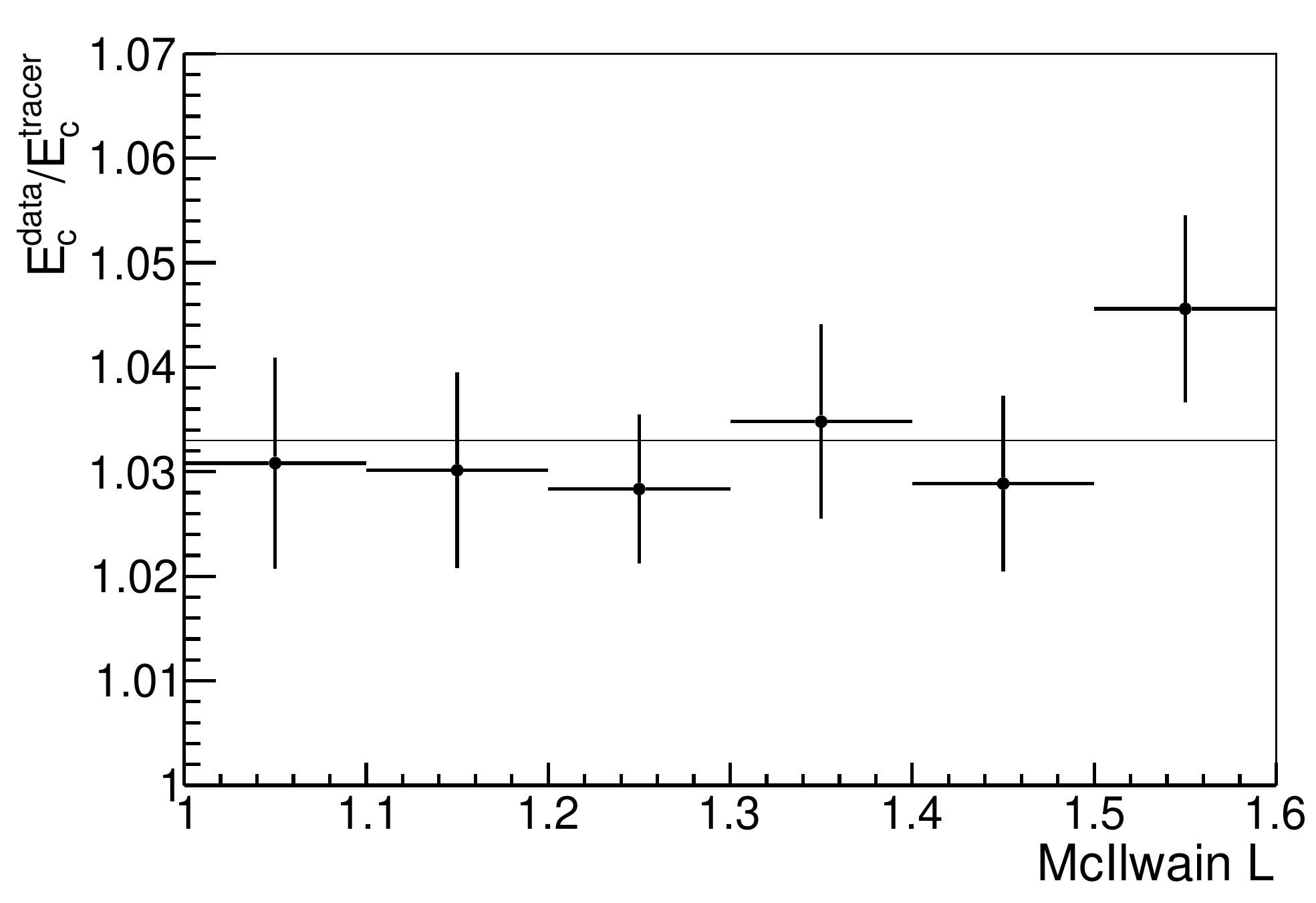}
  \caption{The $E_{c}^{\mathrm{data}}/E_{c}^{\mathrm{tracer}}$ ratio as a function of McIlwain~L. The horizontal line corresponds to the average ratio.}
  \label{fig:absoluteenergyscale_result}
\end{figure}

The $E_{c}^{\mathrm{data}}/E_{c}^{\mathrm{tracer}}$ ratios in the 6 McIlwain~L intervals are in agreement, as shown in Fig.~\ref{fig:absoluteenergyscale_result}, and the average ratio is $1.033 \pm 0.004$. We varied the parameters of the analysis (event selection, energy interval used to derive the template of secondaries and spectral index of the \emph{tracer} spectrum) and found that the average ratio did not vary by more than 0.3\%. Using the IGRF 1995 model changed the result by less than 0.1\%. We performed this analysis in various time periods and the ratio was constant within less than 1\%. We also changed the solar modulation parameter $\phi$ to 0 and 1000~MV and the ratio changed by 0.5\%. We thus estimate the systematic uncertainty of the ratio measurement to be 2\%. The previous LAT measurement of the absolute energy scale found an average ratio of $1.025 \pm 0.005 (\mathrm{stat}) \pm 0.025 (\mathrm{syst})$, which is compatible with the new result.

\section*{Appendix C: Systematic uncertainties as nuisance parameters}


The potential systematic bias in the CRE flux measurement induced by the systematic uncertainties on background subtraction and the IVC corrections can be modeled as a sequence of unknown correction factors. They correspond to nuisance parameters that are derived in the fit of the CRE spectrum when we test different parameterizations for the CRE spectrum.

We define $\mathcal{S}(E)$ as the quadratic sum of the contamination and IVC corrections systematic uncertainties (discussed in section~\ref{sec:systematics}) as a function of energy. We choose $\mathcal{N}$ reference energies $\mathcal{E}_{j}$, logarithmically spaced between 42~GeV and 2~TeV, in order to define $s(E;\mathbf{w})$, a piecewise function, linear in $\log_{10}E$, defined by its values $w_j$ at $\mathcal{E}_{j}$. The set of $w_j$ are the nuisance parameters. The correction factor for the predicted number of counts in the analysis energy bin $i$ is $1+s(E_i;\mathbf{w})\mathcal{S}(E_i)$ and the $\chi^2$ function for the spectral fit is given by:
\begin{equation}
\chi^2 = \sum_{i=1}^{n} \left(\frac{N_{i} - [1+s(E_{i};\mathbf{w})\mathcal{S}(E_{i})] \mu_{i}(\boldsymbol{\theta})}{\delta N_{i}} \right)^2 + \sum_{j=1}^{\mathcal{N}} w_{j}^2
\label{eq:chi2}
\end{equation}
where $\boldsymbol{\theta}$ are the free parameters of the CRE spectral model, $n$ is the number of energy bins of the analysis, $N_{i}$ is the number of counts measured in bin $i$, $\mu_{i}$ is the predicted number of counts after convolution with the DRM and $\delta N_{i}$ is the quadratic sum of the statistical and acceptance uncertainties. The second term of the $\chi^2$ function corresponds to a Gaussian prior on the amplitude of the nuisance parameters.  

The choice of the number of nuisance parameters $\mathcal{N}$ cannot be inferred from first principles. Between the contamination and the IVC corrections systematic uncertainties, the latter dominates. Because we build eight BDTs, the importance of the BDT input variables can change from bin to bin. As a consequence, a change of the IVC correction for one observable can have a significantly different impact even on two adjacent BDT energy bins.

Ignoring any correlation between BDT energy bins would lead to the choice $\mathcal{N} = 8$. But the importance of the input variables depends on the variation of the event topology with energy, which is not expected to change abruptly at the BDT energy boundaries. And this argument also applies for the contamination uncertainty. In order to take into account the BDT bin-to-bin correlation, we chose $\mathcal{N} = 6$. We checked that the fit results do not change significantly with $\mathcal{N} = 5$~or~7.

\begin{figure}[!htb]
  \includegraphics[width=.8\linewidth]{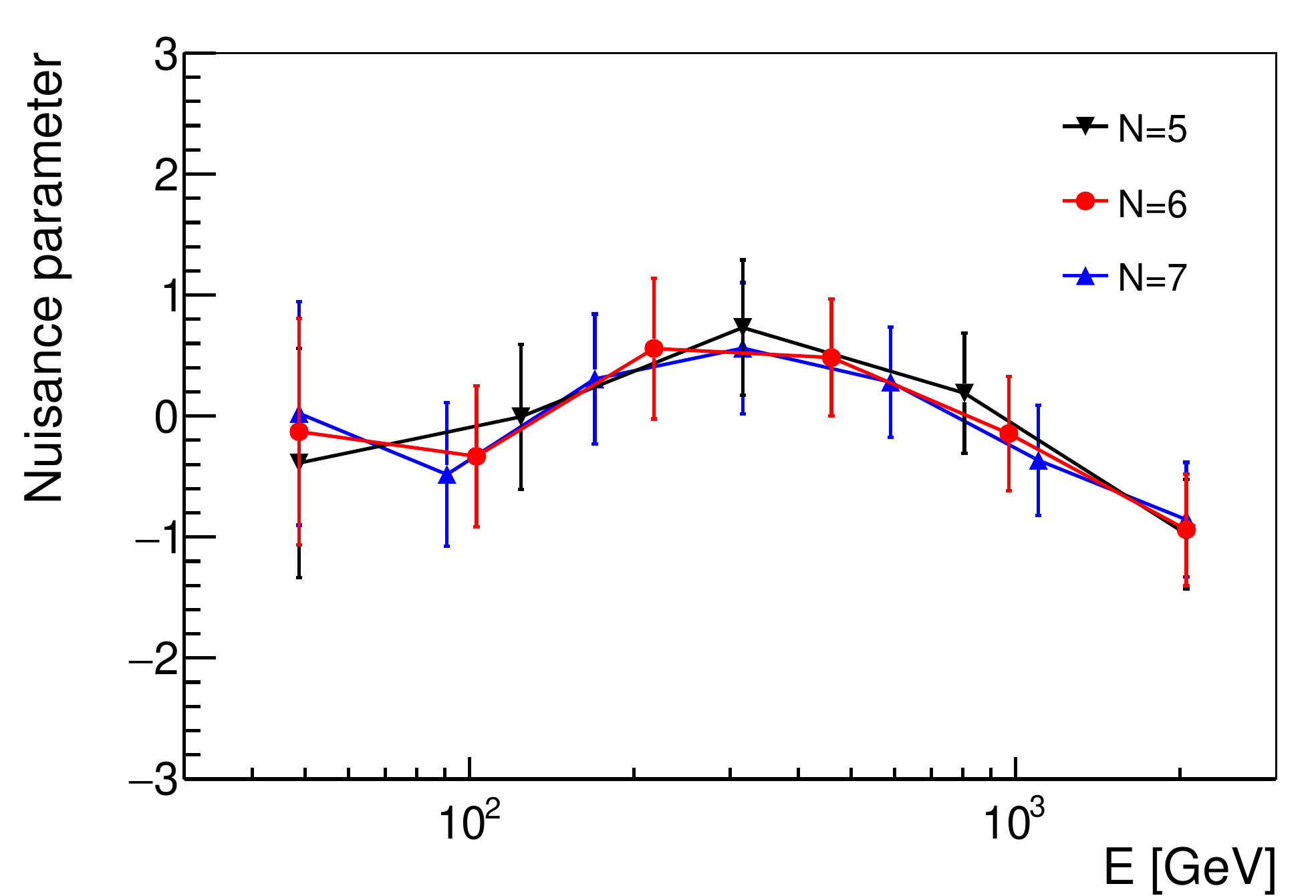}
  \caption{Nuisance parameters found by the fit of the CRE spectrum between 50~GeV and 2~TeV for various numbers of reference energies: $\mathcal{N} = 5$ (black), 6 (red) and 7 (blue).}
  \label{fig:nuisance}
\end{figure}

When fitting the CRE spectrum between 50~GeV and 2~TeV with a single power law and $\mathcal{N} = 6$, we find $\chi^2 = 15.5$ for 18 degrees of freedom. The nuisance parameters values found by the fitting procedure are shown in Fig.~\ref{fig:nuisance}, as well as those found with $\mathcal{N} = 5$~and~7. In all three cases, the nuisance parameters are within $\pm1$ and the spectral index is $3.07 \pm 0.02$.

When fitting with an exponentially cutoff power law $E^{-\gamma} e^{-E/E_c}$, the 95\% CL lower limit on the energy cutoff for $\mathcal{N} = 5$, 6~and~7 is 2.18, 2.13 and 2.19, respectively.

\section*{Appendix D: Tables}

Tables~\ref{tab:table1} and~\ref{tab:table2} give the number of events, the residual background contamination and the CRE flux in each energy bin for the LE and HE analyses, respectively. These tables are available in machine-readable format at \url{https://www-glast.stanford.edu/pub_data/1144}.

\begin{table}[ht]
\centering
\setlength{\tabcolsep}{1pt}
  \begin{tabular}{c c c c}
    \hline
    \hline
  Energy(GeV) & Counts & Cont. & $J_E$(GeV$^{-1}$s$^{-1}$m$^{-2}$sr$^{-1}$)\\
    \hline
7.0-7.8 & 8231 & 0.02 & 5.93 $\times$ (1$\pm$0.016$\pm$0.032) $\times$ 10$^{-01}$\\
7.8-8.7 & 35912 & 0.02 & 4.18 $\times$ (1$\pm$0.008$\pm$0.034) $\times$ 10$^{-01}$\\
8.7-9.7 & 51417 & 0.02 & 3.02 $\times$ (1$\pm$0.007$\pm$0.035) $\times$ 10$^{-01}$\\
9.7-10.8 & 60019 & 0.03 & 2.15 $\times$ (1$\pm$0.007$\pm$0.035) $\times$10$^{-01}$\\
10.8-12.0 & 66545 & 0.03 & 1.53 $\times$ (1$\pm$0.007$\pm$0.036) $\times$ 10$^{-01}$\\
12.0-13.3 & 74725 & 0.02 & 1.09 $\times$ (1$\pm$0.007$\pm$0.030) $\times$ 10$^{-01}$\\
13.3-14.8 & 84090 & 0.03 & 7.90 $\times$ (1$\pm$0.007$\pm$0.029) $\times$ 10$^{-02}$\\
14.8-16.4 & 89109 & 0.03 & 5.67 $\times$ (1$\pm$0.007$\pm$0.030) $\times$ 10$^{-02}$\\
16.4-18.2 & 81203 & 0.03 & 4.02 $\times$ (1$\pm$0.007$\pm$0.030) $\times$ 10$^{-02}$\\
18.2-20.2 & 68111 & 0.03 & 2.88 $\times$ (1$\pm$0.008$\pm$0.029) $\times$ 10$^{-02}$\\
20.2-22.4 & 56832 & 0.03 & 2.06 $\times$ (1$\pm$0.008$\pm$0.028) $\times$ 10$^{-02}$\\
22.4-24.8 & 46535 & 0.03 & 1.47 $\times$ (1$\pm$0.009$\pm$0.022) $\times$ 10$^{-02}$\\
24.8-27.6 & 38267 & 0.03 & 1.07 $\times$ (1$\pm$0.010$\pm$0.017) $\times$ 10$^{-02}$\\
27.6-30.7 & 30449 & 0.03 & 7.54 $\times$ (1$\pm$0.011$\pm$0.018) $\times$ 10$^{-03}$\\
30.7-34.1 & 23408 & 0.04 & 5.33 $\times$ (1$\pm$0.013$\pm$0.020) $\times$ 10$^{-03}$\\
34.1-37.9 & 18867 & 0.04 & 3.87 $\times$ (1$\pm$0.014$\pm$0.019) $\times$ 10$^{-03}$\\
37.9-42.2 & 14718 & 0.05 & 2.72 $\times$ (1$\pm$0.016$\pm$0.017) $\times$ 10$^{-03}$\\
42.2-47.0 & 11186 & 0.05 & 1.92 $\times$ (1$\pm$0.019$\pm$0.018) $\times$ 10$^{-03}$\\
47.0-52.3 & 8618 & 0.06 & 1.35$\times$ (1$\pm$0.021$\pm$0.015) $\times$ 10$^{-03}$\\
52.3-58.5 & 6942 & 0.05 & 9.89 $\times$ (1$\pm$0.023$\pm$0.015) $\times$ 10$^{-04}$\\
58.5-65.3 & 5165 & 0.05 & 6.89 $\times$ (1$\pm$0.027$\pm$0.024) $\times$ 10$^{-04}$\\
65.3-73.0 & 3891 & 0.08 & 4.82 $\times$ (1$\pm$0.034$\pm$0.038) $\times$ 10$^{-04}$\\
\hline
\hline
  \end{tabular}
  \caption{Number of events after background subtraction (without correction for the loss of CREs above the geomagnetic energy cutoff), residual background contamination and flux $J_E$, with its statistical and systematic errors, for the LE selection.}
  \label{tab:table1}
\end{table}


\begin{table}[htb]
\centering
\setlength{\tabcolsep}{2pt}
  \begin{tabular}{c c c c}
    \hline
    \hline
  Energy(GeV) & Counts & Cont. & $J_E$(GeV$^{-1}$s$^{-1}$m$^{-2}$sr$^{-1}$)\\
    \hline
42.2-48.7 & 3948132 & 0.02 & 1.78 $\times$ (1$\pm$0.001$\pm$0.012$\pm$0.013) $\times$ 10$^{-03}$\\
 48.7-56.2 & 2945632 & 0.02 & 1.14 $\times$ (1$\pm$0.001$\pm$0.008$\pm$0.014) $\times$ 10$^{-03}$\\
 56.2-64.9 & 2189648 & 0.02 & 7.42 $\times$ (1$\pm$0.002$\pm$0.027$\pm$0.016) $\times$ 10$^{-04}$\\
 64.9-75.0 & 1609640 & 0.02 & 4.66 $\times$ (1$\pm$0.001$\pm$0.016$\pm$0.017) $\times$ 10$^{-04}$\\
 75.0-86.6 & 1161424 & 0.03 & 2.98 $\times$ (1$\pm$0.002$\pm$0.009$\pm$0.020) $\times$ 10$^{-04}$\\
 86.6-100.0 & 865855 & 0.03 & 1.93 $\times$ (1$\pm$0.002$\pm$0.012$\pm$0.022) $\times$ 10$^{-04}$\\
 100.0-115.5 & 629884 & 0.03 & 1.24 $\times$ (1$\pm$0.002$\pm$0.012$\pm$0.023) $\times$ 10$^{-04}$\\
 115.5-133.4 & 466148 & 0.03 & 8.06 $\times$ (1$\pm$0.002$\pm$0.016$\pm$0.025) $\times$ 10$^{-05}$\\
 133.4-154.0 & 343066 & 0.04 & 5.24 $\times$ (1$\pm$0.002$\pm$0.021$\pm$0.029) $\times$ 10$^{-05}$\\
 154.0-177.8 & 253798 & 0.04 & 3.33 $\times$ (1$\pm$0.003$\pm$0.013$\pm$0.034) $\times$ 10$^{-05}$\\
 177.8-205.4 & 187997 & 0.04 & 2.16 $\times$ (1$\pm$0.003$\pm$0.015$\pm$0.029) $\times$ 10$^{-05}$\\
 205.4-237.1 & 138234 & 0.05 & 1.40 $\times$ (1$\pm$0.003$\pm$0.020$\pm$0.032) $\times$ 10$^{-05}$\\
 237.1-273.8 & 101444 & 0.05 & 8.80 $\times$ (1$\pm$0.004$\pm$0.016$\pm$0.040) $\times$ 10$^{-06}$\\
 273.8-316.2 & 75547 & 0.06 & 5.88 $\times$ (1$\pm$0.005$\pm$0.014$\pm$0.045) $\times$ 10$^{-06}$\\
 316.2-365.2 & 54462 & 0.06 & 3.70 $\times$ (1$\pm$0.005$\pm$0.018$\pm$0.044) $\times$ 10$^{-06}$\\
 365.2-421.7 & 37883 & 0.07 & 2.33 $\times$ (1$\pm$0.006$\pm$0.019$\pm$0.054) $\times$ 10$^{-06}$\\
 421.7-487.0 & 28142 & 0.07 & 1.56 $\times$ (1$\pm$0.007$\pm$0.007$\pm$0.060) $\times$ 10$^{-06}$\\
 487.0-562.3 & 19641 & 0.08 & 9.62 $\times$ (1$\pm$0.008$\pm$0.016$\pm$0.071) $\times$ 10$^{-07}$\\
 562.3-649.4 & 14000 & 0.07 & 6.16 $\times$ (1$\pm$0.009$\pm$0.033$\pm$0.067) $\times$ 10$^{-07}$\\
 649.4-749.9 & 10240 & 0.06 & 4.11 $\times$ (1$\pm$0.010$\pm$0.042$\pm$0.070) $\times$ 10$^{-07}$\\
 749.9-866.0 & 7338 & 0.08 & 2.67 $\times$ (1$\pm$0.012$\pm$0.024$\pm$0.082) $\times$ 10$^{-07}$\\
 866.0-1000.0 & 4938 & 0.10 & 1.63 $\times$ (1$\pm$0.015$\pm$0.024$\pm$0.094) $\times$ 10$^{-07}$\\
 1000.0-1154.8 & 3406 & 0.11 & 1.03 $\times$ (1$\pm$0.018$\pm$0.028$\pm$0.088) $\times$ 10$^{-07}$\\
 1154.8-1333.5 & 2249 & 0.15 & 6.31 $\times$ (1$\pm$0.023$\pm$0.016$\pm$0.097) $\times$ 10$^{-08}$\\
 1333.5-1539.9 & 1491 & 0.13 & 3.83 $\times$ (1$\pm$0.027$\pm$0.075$\pm$0.107) $\times$ 10$^{-08}$\\
 1539.9-1778.3 & 1086 & 0.19 & 2.57 $\times$ (1$\pm$0.036$\pm$0.047$\pm$0.143) $\times$ 10$^{-08}$\\
 1778.3-2053.5 & 737 & 0.22 & 1.62 $\times$ (1$\pm$0.039$\pm$0.077$\pm$0.115) $\times$ 10$^{-08}$\\
    \hline
    \hline
  \end{tabular}
  \caption{Number of events after background subtraction, residual background contamination and flux $J_E$, with its statistical and systematic errors (the acceptance uncertainty and the sum of the contamination and IVC correction uncertainties are shown separately), for the HE selection.}
  \label{tab:table2}
\end{table}

\clearpage

%

\end{document}